\documentclass[aps,pra,showkeys,twocolumn,superscriptaddress]{revtex4-2}
\usepackage{array}
\usepackage{booktabs}
\usepackage{tabu}
\usepackage{dcolumn}
\usepackage{amsmath}
\usepackage{amsfonts}
\usepackage{float}
\usepackage{amssymb}
\usepackage{graphicx,color}
\usepackage[colorlinks={true}]{hyperref}
\hypersetup{colorlinks=true,linkcolor=cyan,citecolor=blue,urlcolor=blue}
\usepackage{graphicx}
\usepackage{subfigure}
\usepackage{graphicx}
\usepackage{dcolumn}
\usepackage{bm}
\usepackage{pstricks}
\usepackage{braket}
\usepackage{orcidlink}

\def\be{\begin{equation}}
  \def\ee{\end{equation}}
\def\bea{\begin{eqnarray}}
\def\eea{\end{eqnarray}}
\def\f{\frac}
\def\n{\nonumber}
\def\l{\label}
\def\p{\phi}
\def\o{\over}
\def\R{\rho}
\def\pa{\partial}
\def\om{\omega}
\def\na{\nabla}
\def\P{\Phi}
\begin{document}

\title{Quantumness near a Schwarzschild black hole}

\author{S. Haddadi \orcidlink{0000-0002-1596-0763}}\email{haddadi@semnan.ac.ir}
\affiliation{Faculty of Physics, Semnan University, P.O. Box 35195-363, Semnan, Iran}

\author{M. A. Yurischev \orcidlink{0000-0003-1719-3884}}
\affiliation{Federal Research Center of Problems of Chemical Physics and Medicinal Chemistry, Russian Academy of Sciences, Chernogolovka 142432, Moscow Region, Russia}

\author{M. Y. Abd-Rabbou \orcidlink{0000-0003-3197-4724}}
\affiliation{Mathematics Department, Faculty of Science, Al-Azhar University, Nasr City 11884, Cairo, Egypt}

\author{M. Azizi \orcidlink{0000-0002-1935-5923}}
\affiliation{Faculty of Physics, Semnan University, P.O. Box 35195-363, Semnan, Iran}

\author{M. R. Pourkarimi \orcidlink{0000-0002-8554-1396}}
\affiliation{Department of Physics, Salman Farsi University of Kazerun, Kazerun, Iran}

\author{M. Ghominejad \orcidlink{0000-0002-0136-7838}}\email{mghominejad@semnan.ac.ir}
\affiliation{Faculty of Physics, Semnan University, P.O. Box 35195-363, Semnan, Iran}

\date{\today}
\def\be{\begin{equation}}
  \def\ee{\end{equation}}
\def\bea{\begin{eqnarray}}
\def\eea{\end{eqnarray}}
\def\f{\frac}
\def\n{\nonumber}
\def\l{\label}
\def\p{\phi}
\def\o{\over}
\def\R{\rho}
\def\pa{\partial}
\def\om{\omega}
\def\na{\nabla}
\def\P{$\Phi$}

\begin{abstract}
The merging of quantum information science with the relativity theory presents novel opportunities for understanding the enigmas surrounding the transmission of information in relation to black holes. For this purpose, we study the quantumness near a Schwarzschild black hole in a practical model under decoherence. The scenario we consider in this paper is that a stationary particle in the flat region interacts with its surroundings while another particle experiences free fall in the vicinity of a Schwarzschild black hole's event horizon. We explore the impacts of Hawking radiation and decoherence on the system under investigation and find that these effects can limit the survival of quantum characteristics, but cannot destroy them completely. Hence, the results of this study possess the potential to yield valuable insights into the comprehension of the quantum properties of a real system operating within a curved space-time framework.
\end{abstract}


\maketitle

\section{INTRODUCTION}	
In 2000, Zurek introduced the concept of information-theoretic quantum discord between two classically identical definitions of mutual information \cite{Zurk2000}. Quantum discord can be regarded as a measure of a violation of the classicality of a joint state of two quantum subsystems. In the next year, he, alongside Ollivier, incorporated the optimization procedure into his original definition of quantum discord \cite{Zurk2001}. Therefore, for the first time in their paper, we encountered the term ``quantumness" of correlations.
Quantum correlations \cite{Henderson2001}, a fundamental aspect of quantum mechanics, are crucial in numerous quantum information processing tasks, including quantum cryptography, quantum communication, and quantum computation \cite{Nielsen2000}. Nonetheless, these correlations are susceptible to fragility owing to the effect of a variety of decoherence \cite{Breuer2002}. Literally, decoherence is the phenomenon wherein a quantum system interacts with its environment, resulting in the loss of quantum coherence and quantum correlations \cite{Rivas2012}. Generally, we know that quantum systems interact with their environment in the real world \cite{mlhu2018}. As a result, quantum correlations present in the initial state may experience a decay over time. In the field of quantum computing, the computation process is susceptible to accumulating errors caused by various decoherence mechanisms. These errors can disrupt the accuracy of quantum gates and ultimately influence the precision of the final results. Unfortunately, such errors have the potential to affect the correlations necessary for quantum algorithms to function properly.

On the one hand, environmental noise, which includes various external factors such as electromagnetic radiation and temperature fluctuations, has the potential to cause stochastic fluctuations in the parameters of a considered quantum system \cite{Kraus1983}. These fluctuations possess the potential to induce uncontrolled phase shifts and energy exchanges, thereby disturbing the quantum correlations. On the other hand, dephasing occurs when different parts of a quantum system's wave function acquire different phases owing to interactions with surrounding \cite{Cohen1999}. The phenomenon of dephasing can lead to the dissipation of quantum correlation and coherence between quantum states \cite{Hu2019,Hu2021}. The correlations that are most vulnerable to dephasing are those that pertain to superpositions and phase differences, such as those that exist in Bell states utilized for quantum entanglement. Moreover, the spontaneous emission may lead to a loss of coherence in quantum states and result in reduced correlations, particularly in cases where photon emission affects the entangled states \cite{Tanas2004}.

In general, quantum measurements inherently perturb the measured quantum state \cite{me1,me2,me3}. This disturbance can result in the collapse of the state into an eigenstate of the measured observable, which can affect the correlations \cite{me4}. Repeated measurements can also perturb the correlations over time, especially where entanglement is involved. However, in order to reduce the destructive effects of decoherence on quantum correlations, various techniques such as error correction codes, quantum error correction protocols, and decoherence-free subspaces \cite{Ekert2000} are employed. These approaches help protect and extend the lifetime of quantum correlations in the presence of environmental influences \cite{Khalid2020,Czerwinski2022}.

The dynamics of quantum correlations under decoherence effects and Hawking radiation can be quite complex but intriguing, involving an intricate interplay between gravity, quantum information, and environmental interactions \cite{Fuentes2005,Alsing2006,Martin2010,Bruschi2010,Mann2012,Ahmadi2016,Garcia2019}. Hawking radiation appears near the event horizon of a black hole due to quantum effects in the vicinity of the event horizon boundary \cite{HAWKING1974}. Particles are created in pairs near the event horizon, where one particle falls into the black hole and the other escapes as radiation. This process causes black holes to gradually lose their mass over time. Besides, Hawking radiation can be witnessed as a form of decoherence because of the interaction of quantum fields near the black hole. This process can introduce disturbances and thermal noise that can affect quantum correlations, just like environmental decoherence.

Interestingly, the process of Hawking radiation may generate entangled pairs of particles near the black hole's event horizon \cite{Pan2008,WangPan2009,Iizuka2013}. These produced entangled pairs can carry away some of the black hole's energy, contributing to the mass loss of the black holes. Potentially, Hawking radiation can also carry away entanglement, affecting the correlations between the infalling and escaping particles. Notably, the interplay between quantum correlations and Hawking radiation is linked to the black hole information paradox. Although the information is conserved according to quantum mechanics, the Hawking process seems to present that information can be lost when particles fall into a black hole. This has led to debates regarding whether and how quantum correlations can be reconstructed or preserved in the presence of Hawking radiation.

Studying the combined effects of Hawking radiation and decoherence requires a deep understanding of both quantum field theory and quantum information theory, as well as a grasp of the underlying principles of black holes and gravity.  While some theoretical progress has been made in exploring these interactions  \cite{xw01,xw02,xw02b,xw03,xw04,xw05,xw06,xw07,Zhangadp2018,xw08,xw09,Abd-RabbouIJQI,Abd-RabbouScr,xw10,xw11,xw12,xw13,xw14,xw15} and the quantumness of various systems \cite{DW01,DW02,DW03,DW04}, many aspects are still the subject of ongoing research and debate. With this in mind, we are motivated to explore the quantumness (quantum correlations, quantum coherence, and non-locality) of a physical system consisting of two qubits in the background of the Schwarzschild black hole under the influence of both the decoherence effect and Hawking radiation.

\section{Measures of quantum information correlation}\label{sec:2}

\subsection{Local quantum uncertainty}
The local quantum uncertainty (LQU) is used when dealing with measurements and interactions at a specific location or in a limited region of space. It is a foundational characteristic of quantum mechanics that challenges our classical intuition and also underlies many of the unique behaviors and phenomena observed in the quantum world. Indeed, it is a measure of quantum correlations based on the Wigner-Yanase skew information $\mathcal{I}$ \cite{wigner1963,luo2003}. The LQU pertaining to subspace of party $A$ with the measurement operator $H_A$, upon optimization over all local observables on $A$, is explicitly given as \cite{Gilorami2013}
\begin{equation}\label{lqu1}
\mathcal{U}(\varrho):=\min _{H_A} \mathcal{I}\left(\varrho, H_A\right).
\end{equation}

Therefore, the concept of LQU is precisely characterized as the minimum level of quantum uncertainty that is inherently linked to a singular measurement pertaining to one of the subsystems of the bipartite system, $AB$. It is noteworthy to mention that LQU is a genuine measure of quantum correlations, and it has been demonstrated that LQU satisfies all of the requisite physical conditions that are necessary to qualify as a criterion of quantum correlations. Gilorami \textit{et al.} \cite{Gilorami2013} were able to execute optimization for qubit-qudit systems successfully and subsequently introduced the measure \eqref{lqu1} in a specific form as
\begin{equation}\label{lqu2}
\mathcal{U}=1-\lambda_{\max}^{\mathcal{W}},
\end{equation}
where $\lambda_{\max}^{\mathcal{W}}$ represents the highest eigenvalue of the three-by-three symmetric matrix $\mathcal{W}$, which comprises entries
\begin{equation}
\mathcal{W}_{\nu \mu}=\textmd{tr}\{\sqrt{\varrho}(\sigma_{\nu}\otimes I)\sqrt{\varrho}(\sigma_{\mu}\otimes I)\},
\end{equation}
where $\sigma_{\nu}$ and $\sigma_{\mu}$ represent the set of Pauli matrices with $\nu, \mu=x, y, z$.

\subsection{Local quantum Fisher information}
Quantum Fisher information (QFI) is a concept from quantum metrology, which deals with the precision of measurements in the quantum realm \cite{qfi1,qfi2,qfi3,qfi5}. In quantum mechanics, the Fisher information quantifies how much information about a parameter of interest is contained in the outcomes of measurements. It provides a measure of how well a quantum state can be distinguished from nearby states in terms of the parameter being measured \cite{qfi7}. The local quantum Fisher information (LQFI) measure $\mathcal{F}$ is based on the QFI (indicated by $F$) and it is equivalent to the optimal LQFI when the measurement operator $H_A$ is applied to the subsystem $A$ within the bipartite system $AB$. Thus, the LQFI is as follows \cite{qfi7}
\begin{equation}
\mathcal{F}(\varrho):=\min _{H_A} F\left(\varrho, H_A\right).
\end{equation}

Specifically, one can write the following formula for LQFI if the subsystem $A$ is a qubit
\begin{equation}\label{lqfi2}
\mathcal{F}=1-\lambda_{\max}^{\mathcal{M}},
\end{equation}
where $\lambda_{\max}^{\mathcal{M}}$ denotes the highest eigenvalue of the 3-by-3 symmetric matrix $\mathcal{M}$, which its elements are
\begin{equation}
\mathcal{M}_{\nu \mu}=\sum_{i,j; q_i + q_j \neq 0} \frac{2 q_i q_j}{q_i + q_j} \langle \psi_i |\sigma_\nu \otimes I|\psi_j\rangle \langle \psi_j |\sigma_\mu \otimes I|\psi_i\rangle,
\end{equation}
where $q_i$ and $|\psi_i\rangle$ are respectively the eigenvalues and eigenstates of $\varrho=\Sigma_i q_i |\psi_i\rangle \langle\psi_i|$ with $q_i \geq 0$ and $\Sigma_i q_i=1$.

\section{Theoretical framework}\label{sec:3}
Let us commence by revisiting the precise definition of metric Hawking radiation as it pertains to the Dirac field model in Schwarzschild black hole. Typically, the mathematical model known as the Dirac equation is utilized to explain the behavior of particles in relation to a space-time that is curved in a general manner. This phenomenon is commonly expressed through the following formula \cite{Brill1957,Jing2004}
\begin{equation}
\left[\gamma^a e_a^\mu\left(\partial_\mu+\Gamma_\mu\right)\right] \Psi=0,
\end{equation}
where $\gamma^a$ is Dirac matrix, $\Gamma_\mu$ is spin connection, $e_a^\mu$ is the inverse of tetrad $e_\mu^a$, and $\mu$ is mass of the Dirac field. In the background of Schwarzschild, the metric may be delineated as a mathematical construct that characterizes the spatial and temporal properties of space-time. This metric can be defined as
\begin{equation}
d s^2=-f d t^2+\frac{1}{f} d r^2+r^2\left(d \theta^2+\sin ^2 \theta d \varphi^2\right),
\end{equation}
with $f=1-2Mr^{-1}$, where $M$ denotes the mass of the black hole and $r$ is the radial coordinates. In this paper, the gravitational constant, reduced Planck constant, speed of light, and Boltzmann constant are assumed to be equal to one for simplicity.

Within the framework of Schwarzschild space-time, it is now possible to derive the Dirac equation as
\begin{equation}
\begin{aligned}\label{Dirac equation}
& -\gamma_0 f^{-1/2} \frac{\partial \Psi}{\partial t}+\gamma_1 f^{1/2} \left[\frac{\partial}{\partial r}+r^{-1}+\frac{f^{-1}M}{2 r^2}\right] \Psi \\
& +\frac{\gamma_2}{r} \left(\frac{\partial }{\partial \theta}+\frac{\cot \theta}{2}\right)\Psi+\frac{\gamma_3}{r \sin \theta} \frac{\partial \Psi}{\partial \varphi}=0.
\end{aligned}
\end{equation}

By solving equation \eqref{Dirac equation}, it is feasible to derive the outgoing solutions of positive (fermions) frequency for both the outside and inside domains surrounding the event horizon as \cite{Jing2004}
\begin{equation}
\Psi_{k,out}^{\mathrm{I}+}=\mathcal{S} e^{-i \omega u},
\end{equation}
and
\begin{equation}
\Psi_{k,in}^{\mathrm{II}+}=\mathcal{S} e^{i \omega u},
\end{equation}
where $\mathcal{S}$ is a 4-component Dirac spinor, $k$ is a wave vector, $\omega$ is a monochromatic frequency of Dirac filed, and $u$ is the retarded time expressed as $u=t-r^*$  with $r^{*}=r+2M\ln[rf/2M]$ which is the tortoise coordinate.

By employing Damour and Ruffini's suggestion \cite{Damoar1976}, an analytic extension can be made for the aforementioned equation, thereby providing a comprehensive foundation for the positive energy modes. As a result, it becomes possible to obtain the Bogoliubov transformations \cite{Bogoljubov1958,Radmore1997} that pertain to the creation and annihilation operators in both the Schwarzschild and Kruskal coordinates by quantizing the Dirac fields in the Schwarzschild and Kruskal modes, respectively. Upon suitably normalizing the state vector, one can express the expressions of the Kruskal vacuum and excited states with mode $k$ as
\begin{align}
|0\rangle_k^{+} & \rightarrow \alpha\left|0_k\right\rangle_{\mathrm{I}}^{+}\left|0_{-k}\right\rangle_{\mathrm{II}}^{-}+\beta\left|1_k\right\rangle_{\mathrm{I}}^{+}\left|1_{-k}\right\rangle_{\mathrm{II}}^{-}, \nonumber\\
|1\rangle_k^{+} & \rightarrow\left|1_k\right\rangle_{\mathrm{I}}^{+}\left|0_{-k}\right\rangle_{\mathrm{II}}^{-},\label{eq13}
\end{align}
where $\alpha=(e^{-\omega_k / T}+1)^{-1/2}$ and $ \beta=(e^{\omega_k / T}+1)^{-1/2}$
 with $T=1/8\pi M$ which is the Hawking temperature \cite{Kerner2006}. Besides, $\left|n_k\right\rangle_{\mathrm{I}}^{+}$ and $\left|n_{-k}\right\rangle_{\mathrm{II}}^{-}$ are respectively represented as the orthonormal bases for outside and inside domains of the event horizon. For the sake of simplicity, it is imposed that $\omega_k=\omega=1$, $\left|n_k\right\rangle_{\mathrm{I}}^{+}=\left|n\right\rangle_{\mathrm{I}}$, and $\left|n_{-k}\right\rangle_{\mathrm{II}}^{-}=\left|n\right\rangle_{\mathrm{II}}$.

\begin{figure}[t]
	\begin{center}
     \includegraphics[width=0.5\textwidth]{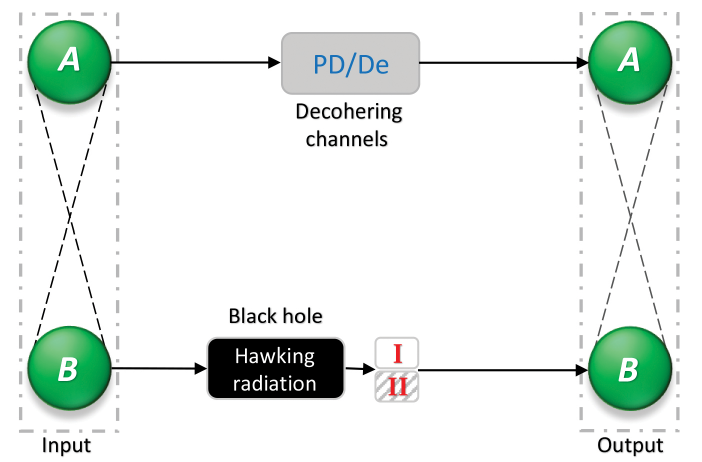}	
	\end{center}
	\caption{A schematic diagram for the physical model with Alice's particle--$A$ in a flat region and Bob's particle--$B$ near the event horizon of the Schwarzschild black hole. The dashed lines indicate the quantum correlation between two particles $A$ and $B$. Input state is given in Eq. \eqref{Bell-diagonal} and output states are provided in Eqs. \eqref{statePD} and \eqref{stateDe}.}
	\label{figure0}
\end{figure}

Finally, in order to investigate the collective effects of Hawking and decohering effects on quantum correlation, we shall consider a system comprising two particles (Alice's particle--$A$ and Bob's particle--$B$) that begin in a state characterized by generic Bell-diagonal state as follows
\begin{equation}\label{Bell-diagonal}
\varrho_{A B}=\frac{1}{4}\left(I_A \otimes I_B+\sum_{i=1}^3 c_i \sigma_i^A \otimes \sigma_i^B\right),
\end{equation}
where $\sigma_1=\sigma_x$, $\sigma_2=\sigma_y$ and $\sigma_3=\sigma_z$, and $c_i=\textmd{tr}_{AB}(\varrho_{A B} \sigma_i^A \otimes \sigma_i^B)$ satisfies $0\leq |c_i| \leq 1$.

In the present study, we suppose that the particle belonging to Bob is situated within a Schwarzschild space-time, which is situated proximal to the event horizon, whilst the particle belonging to Alice is kept within a static and flat space-time. Hence, the state of our considered system \eqref{Bell-diagonal} after interaction of Bob's particle with the Hawking radiation \eqref{eq13} is given by \cite{xw06}
{\small
\begin{equation}
\begin{aligned}
&\varrho_{A B_{\mathrm{I}} B_{\mathrm{II}}}= \\
&\frac{1}{4}\Big[(1+c_3)\{\alpha^2|000\rangle\langle 000|+\alpha \beta(| 000\rangle\langle 011| +|011\rangle\langle 000|)\\
&+\beta^2| 011\rangle\langle 011|+| 110\rangle\langle 110|\}+(1-c_3)\{\alpha^2|100\rangle\langle 100| \\
& +\alpha \beta(| 100\rangle\langle 111|
+|111\rangle\langle 100|)+\beta^2| 111\rangle\langle 111|+| 010\rangle\langle 010|\} \\
& +(c_1-c_2)\{\alpha(|000\rangle\langle 110|+|110\rangle\langle 000|)+\beta(| 011\rangle\langle 110| \\
& +| 110\rangle\langle 011|)\}+(c_1+c_2)\{\alpha(|010\rangle\langle 100|+| 100\rangle\langle 010|) \\
& +\beta(|010\rangle\langle 111|+| 111\rangle\langle 010|)\}\Big]. \label{8by8}
\end{aligned}
\end{equation}}

\section{LQU and LQFI under decoherence}\label{sec:4}
Quantum correlation and coherence are of paramount importance in both quantum information processing and quantum communication protocols. Nonetheless, as stated in the introduction, the interaction of quantum systems with their surroundings triggers a phenomenon commonly referred to as decoherence, which results in the loss of quantum coherence and the disruption of intricate quantum correlations. Consequently, the system may exhibit more classical behavior and forfeit some of its distinctive quantum characteristics.

The physical processes that give rise to decoherence in quantum systems are known as decohering channels. These channels are often depicted as noise or perturbations that impact the quantum state and cause it to become mixed or probabilistic, thereby deviating from its pure state \cite{Nielsen2000,Holevo2019}.
In broad terms, the behavior of particles that independently interact with various environments can be elucidated through the solutions derived from the relevant Born-Markov-Lindblad equations.

In the case of any quantum state that is initially established as $\varrho$, the final quantum state that emerges due to the effects of decohering channels, with the assistance of the Kraus operator approach, can be ascertained by
\begin{equation}
\varepsilon\left(\varrho\right):=\sum_{i} K_{i} \varrho (K_{i})^{\dagger}.\label{Kraus}
\end{equation}

In the above equation, the Kraus operators $K_i$ are denoted as the single-qubit quantum channels. Note that these operators are required to conform to the closure condition, which mandates that the summation of $(K_{i})^{\dagger}K_{i}$ must always be equal to the identity matrix $I$.

In the following, we study the quantum correlation attributes that are associated with our considered system, which is subject to decoherence effects. As is known, there are several decohering channels such as phase damping (PD), depolarizing (De), phase flip (PF), amplitude damping (AD)  and so on \cite{Nielsen2000}. However, in a realistic setting, we consider here two specific channels, namely PD and De, to investigate their different effects on the system in question.

\subsubsection{PD channel}
A PD channel is a quantum communication concept that models a specific type of noise or decoherence that can affect a quantum system during transmission or storage. It represents the loss of information about the phase of a quantum state while preserving its energy. In other words, the amplitude of the quantum state remains unchanged, but its phase information becomes uncertain or randomized due to interactions with the environment. Mathematically, the PD channel can be represented using Kraus operators. For a single-qubit, the Kraus operators of PD channel are
\begin{equation}
K_0=|0\rangle\langle0|+\sqrt{1-p}|1\rangle\langle1|, \quad K_1=\sqrt{p}|1\rangle\langle1|.\label{kpd}
\end{equation}

The channel applies one of these Kraus operators with probabilities determined by $p$ to the input quantum state.

By employing Eqs. \eqref{8by8}, \eqref{Kraus} and \eqref{kpd}, we can obtain the decohered state after passing Alice's particle through the PD decohering channel. Finally, after tracing over all degrees of freedom in the physically inaccessible region (inside the region of the event horizon--$\mathrm{II}$, see Fig. \ref{figure0}), the system's state takes the form
\begin{align}\label{statePD}
\varrho_{A B_{\mathrm{I}}}^{\textmd{PD}}&=
v^{+}|00\rangle\langle00|+\mu^{+}|01\rangle\langle01|+v^{-}|10\rangle\langle10|\nonumber\\
&+\mu^{-}|11\rangle\langle11|+u^{-}|11\rangle\langle00|+u^{+}|10\rangle\langle01|\nonumber\\
&+u^{+}|01\rangle\langle10|+u^{-}|00\rangle\langle11|,
\end{align}
where $v^{\pm}=\alpha^2 (1\pm c_{3})/4$, $\mu^{\pm}=[2-\alpha^2(1\pm c_{3})]/4$, and $u^{\pm}=\alpha\sqrt{1-p} (c_{1}\pm c_{2})/4$.

Using now the general formulas presented in \cite{Yurischev2023}, we derive the analytical expression of LQU \eqref{lqu2} for the above state as follows
\begin{equation}\label{lquPD}
\mathcal{U}(\varrho_{A B_{\mathrm{I}}}^{\textmd{PD}})=1- \max\{\mathcal{W}_{11}^{\textmd{PD}},  \mathcal{W}_{33}^{\textmd{PD}}\},
\end{equation}
where
\begin{align}\label{eq19}
\mathcal{W}_{11}^{\textmd{PD}}=&\left(\sqrt{\gamma_1}+\sqrt{\gamma_2}\right)\left(\sqrt{\gamma_3}+\sqrt{\gamma_4}\right)\nonumber\\
&+\frac{(\mu^{+}-v^-)(\mu^{-}-v^+)+4|u^{-} u^{+}|}{\left(\sqrt{\gamma_1}+\sqrt{\gamma_2}\right)\left(\sqrt{\gamma_3}+\sqrt{\gamma_4}\right)}, \nonumber\\
\mathcal{W}_{33}^{\textmd{PD}}=&\frac{1}{2}\bigg[\left(\sqrt{\gamma_1}+\sqrt{\gamma_2}\right)^2+\left(\sqrt{\gamma_3}+\sqrt{\gamma_4}\right)^2
\nonumber\\
&+\frac{(\mu^{-}-v^{+})^2 - 4|u^{-}|^2}{\left(\sqrt{\gamma_1}+\sqrt{\gamma_2}\right)^2}-\frac{(\mu^{+}-v^{-})^2 - 4|u^{+}|^2}{\left(\sqrt{\gamma_3}+\sqrt{\gamma_4}\right)^2}\bigg],
\end{align}
with $\gamma_{1,2}=(v^{+}+\mu^{-} \pm \sqrt{(v^{+}-\mu^{-})^2 + 4|u^{-}|^2})/2$ and $\gamma_{3,4}=(\mu^{+}+v^{-} \pm \sqrt{(\mu^{+}-v^{-})^2 + 4|u^{+}|^2})/2$.

By defining the branches $\mathcal{U}_0^{\textmd{PD}}=1-\mathcal{W}_{11}^{\textmd{PD}}$ and $\mathcal{U}_1^{\textmd{PD}}=1-\mathcal{W}_{33}^{\textmd{PD}}$, we can rewrite Eq. \eqref{lquPD} as below
\begin{equation}\label{lqubranches}
\mathcal{U}(\varrho_{A B_{\mathrm{I}}}^{\textmd{PD}})=\min\{\mathcal{U}_0^{\textmd{PD}}, \mathcal{U}_1^{\textmd{PD}} \}.
\end{equation}

Besides, utilizing the general equations from \cite{Yurischev2023}, the analytical expression of LQFI \eqref{lqfi2} for the state \eqref{statePD} can be obtained as
\begin{equation}\label{lqfiPD}
\mathcal{F}(\varrho_{A B_{\mathrm{I}}}^{\textmd{PD}})=1- \max\{\mathcal{M}_{11}^{\textmd{PD}},  \mathcal{M}_{33}^{\textmd{PD}}\},
\end{equation}
where
\begin{equation}\label{eq21}
\mathcal{M}_{11}^{\textmd{PD}}=\frac{k_1 k_2}{k_3},\quad \mathcal{M}_{33}^{\textmd{PD}}=1-4\left(\frac{|u^{-}|^2}{v^{+}+\mu^{-}}+\frac{|u^{+}|^2}{\mu^{+}+v^{-}}\right),
\end{equation}
with {\small $$k_1=64\left(v^{+} v^{-}+\mu^{+} \mu^{-}+\gamma_1 \gamma_2+\gamma_3 \gamma_4+2|u^{+} u^{-}|\right),$$ $$k_2=(\mu^{+}+v^{-}) \gamma_1 \gamma_2 +(v^{+}+\mu^{-}) \gamma_3 \gamma_4,$$ and $$k_3=\left[1-\left(\gamma_1-\gamma_2\right)^2-\left(\gamma_3-\gamma_4\right)^2\right]^2-4\left(\gamma_1-\gamma_2\right)^2\left(\gamma_3-\gamma_4\right)^2.$$}

Further, we can get two branches of LQFI as $\mathcal{F}_0^{\textmd{PD}}=1-\mathcal{M}_{11}^{\textmd{PD}}$ and $\mathcal{F}_1^{\textmd{PD}}=1-\mathcal{M}_{33}^{\textmd{PD}}$. Therefore, Eq. \eqref{lqfiPD} is rewritten as
\begin{equation}\label{lqfibranches}
\mathcal{F}(\varrho_{A B_{\mathrm{I}}}^{\textmd{PD}})=\min\{\mathcal{F}_0^{\textmd{PD}}, \mathcal{F}_1^{\textmd{PD}} \}.
\end{equation}

\subsubsection{De channel}
The De channel refers to a type of quantum channel that introduces noise into a quantum system by causing a loss of information about the initial quantum state. This noise tends to depolarize the quantum state, meaning it reduces the coherence and fidelity of the quantum information. A De channel is often represented using a Kraus operator sum representation. For a single-qubit, the Kraus operators of De channel can be written as
\begin{align}
&K_0=\sqrt{1-p} (|0\rangle\langle0|+|1\rangle\langle1|), \, K_1=\sqrt{\frac{p}{3}} (|0\rangle\langle1|+|1\rangle\langle0|) ,\nonumber\\
&K_2= i\sqrt{\frac{p}{3}}(|0\rangle\langle1|-|1\rangle\langle0|), \, K_3= \sqrt{\frac{p}{3}}(|0\rangle\langle0|-|1\rangle\langle1|). \label{kde}
\end{align}

According to the previous method, by exploiting the Eqs. \eqref{8by8}, \eqref{Kraus} and \eqref{kde}, and after tracing over all degrees of freedom in the region $\mathrm{II}$ (see Fig. \ref{figure0}), the final decohered state takes the following form
\begin{align}\label{stateDe}
\varrho_{A B_{\mathrm{I}}}^{\textmd{De}}&=
\vartheta^{+}|00\rangle\langle00|+\eta^{+}|01\rangle\langle01|+\vartheta^{-}|10\rangle\langle10|\nonumber\\
&+\eta^{-}|11\rangle\langle11|+\kappa^{-}|11\rangle\langle00|+\kappa^{+}|10\rangle\langle01|\nonumber\\
&+\kappa^{+}|01\rangle\langle10|+\kappa^{-}|00\rangle\langle11|,
\end{align}
where
$$\vartheta^{\pm}=\frac{1}{12}[2p\mp\alpha^{2}(c_3 \pm 1)(-3+2p)],$$
$$\eta^{\pm}=\frac{1}{12}[6-2p\pm\alpha^{2}(c_3 \pm 1)(-3+2p)],$$
$$\kappa^{\pm}=\frac{1}{12}\alpha[3(c_1 \pm c_2)-2p(c_1\pm 2c_2)].$$

The analytical expression of LQU \eqref{lqu2} for our state \eqref{stateDe} with the branches $\mathcal{U}_0^{\textmd{De}}=1-\mathcal{W}_{11}^{\textmd{De}}$ and $\mathcal{U}_1^{\textmd{De}}=1-\mathcal{W}_{33}^{\textmd{De}}$ can be obtained as
\begin{equation}\label{lquDe}
\mathcal{U}(\varrho_{A B_{\mathrm{I}}}^{\textmd{De}})=\min\{\mathcal{U}_0^{\textmd{De}},  \mathcal{U}_1^{\textmd{De}}\},
\end{equation}
where
\begin{align}\label{eq25}
\mathcal{W}_{11}^{\textmd{De}}=&\left(\sqrt{\theta_1}+\sqrt{\theta_2}\right)\left(\sqrt{\theta_3}+\sqrt{\theta_4}\right)\nonumber\\
&+\frac{(\eta^{+}-\vartheta^-)(\eta^{-}-\vartheta^+)+4|\kappa^{-} \kappa^{+}|}{\left(\sqrt{\theta_1}+\sqrt{\theta_2}\right)\left(\sqrt{\theta_3}+\sqrt{\theta_4}\right)}, \nonumber\\
\mathcal{W}_{33}^{\textmd{De}}=&\frac{1}{2}\bigg[\left(\sqrt{\theta_1}+\sqrt{\theta_2}\right)^2+\left(\sqrt{\theta_3}+\sqrt{\theta_4}\right)^2
\nonumber\\
&+\frac{(\eta^{-}-\vartheta^{+})^2 - 4|\kappa^{-}|^2}{\left(\sqrt{\theta_1}+\sqrt{\theta_2}\right)^2}-\frac{(\eta^{+}-\vartheta^{-})^2 - 4|\kappa^{+}|^2}{\left(\sqrt{\theta_3}+\sqrt{\theta_4}\right)^2}\bigg],
\end{align}
with $\theta_{1,2}=(\vartheta^{+}+\eta^{-} \pm \sqrt{(\vartheta^{+}-\eta^{-})^2 + 4|\kappa^{-}|^2})/2$ and $\theta_{3,4}=(\eta^{+}+\vartheta^{-} \pm \sqrt{(\eta^{+}-\vartheta^{-})^2 + 4|\kappa^{+}|^2})/2$.

Moreover, we obtain the analytical expression of LQFI \eqref{lqfi2} for the mentioned state as
\begin{equation}\label{lqfiDe}
\mathcal{F}(\varrho_{A B_{\mathrm{I}}}^{\textmd{De}})=\min\{\mathcal{F}_0^{\textmd{De}}, \mathcal{F}_1^{\textmd{De}} \}.
\end{equation}
where $\mathcal{F}_0^{\textmd{De}}=1-\mathcal{M}_{11}^{\textmd{De}}$ and $\mathcal{F}_1^{\textmd{De}}=1-\mathcal{M}_{33}^{\textmd{De}}$,
with
\begin{equation}\label{eq27}
\mathcal{M}_{11}^{\textmd{De}}=\frac{l_1 l_2}{l_3},\quad \mathcal{M}_{33}^{\textmd{De}}=1-4\left(\frac{|\kappa^{-}|^2}{\vartheta^{+}+\eta^{-}}+\frac{|\kappa^{+}|^2}{\eta^{+}+\vartheta^{-}}\right),
\end{equation}
and {\small $$l_1=64\left(\vartheta^{+} \vartheta^{-}+\eta^{+} \eta^{-}+\theta_1 \theta_2+\theta_3 \theta_4+2|\kappa^{+} \kappa^{-}|\right),$$ $$l_2=(\eta^{+}+\vartheta^{-}) \theta_1 \theta_2 + (\vartheta^{+}+\eta^{-}) \theta_3 \theta_4,$$  $$l_3=\left[1-\left(\theta_1-\theta_2\right)^2-\left(\theta_3-\theta_4\right)^2\right]^2-4\left(\theta_1-\theta_2\right)^2\left(\theta_3-\theta_4\right)^2.$$}

Notice that the analytical expressions of $\mathcal{W}_{11}$, $\mathcal{W}_{33}$,
$\mathcal{M}_{11}$, and $\mathcal{M}_{33}$ in the cases of high- and low-temperatures for PD and De channels are reported in appendix \ref{Appendix A}.

So, now everything is ready to start a comparative analysis between the LQU and LQFI that were discussed above. The aim is to explore the qualitative as well as quantitative aspects of the quantum correlations in the vicinity of a Schwarzschild black hole when subjected to the effects of decoherence. Without loss of generality, we are motivated to consider an initial maximally non-classically correlated state (pure Bell state) throughout this paper under the condition $c_1=-c_2=c_3=1$ because we are interested in knowing how much quantumness between the non-classically correlated particles of Alice and Bob is lost under the influences of Hawking radiation and decoherence.

\begin{figure*}[t]
	\begin{center}
		\includegraphics[width=0.45\textwidth]{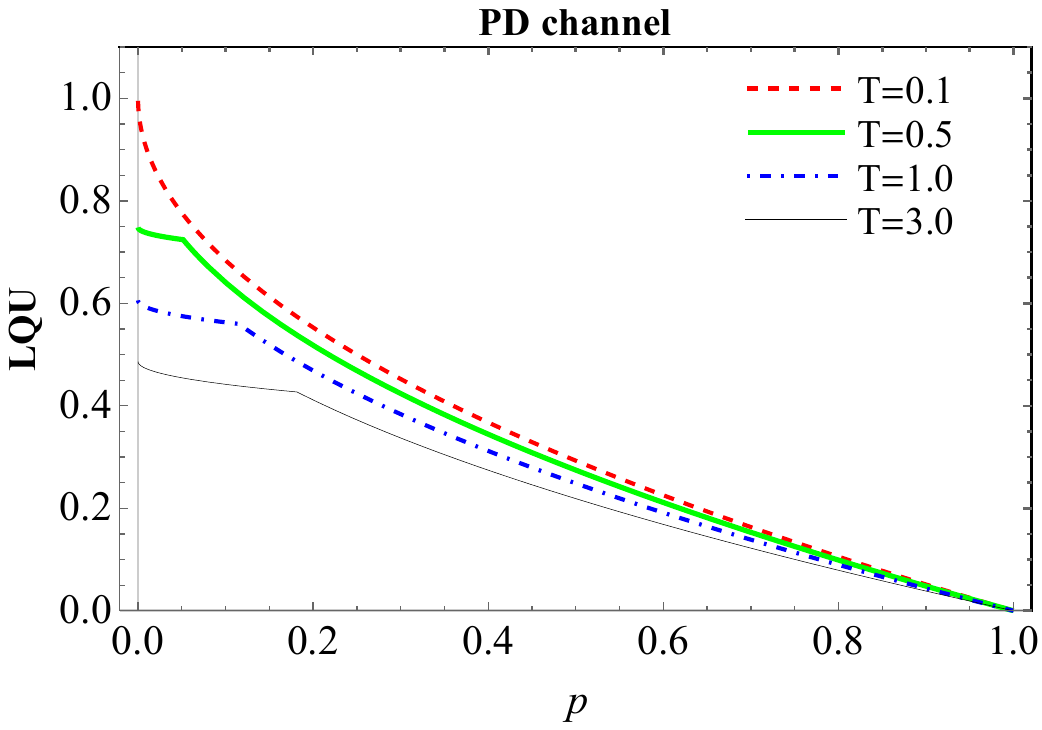}
		\put(-200,155){(a)}\quad
		\includegraphics[width=0.45\textwidth]{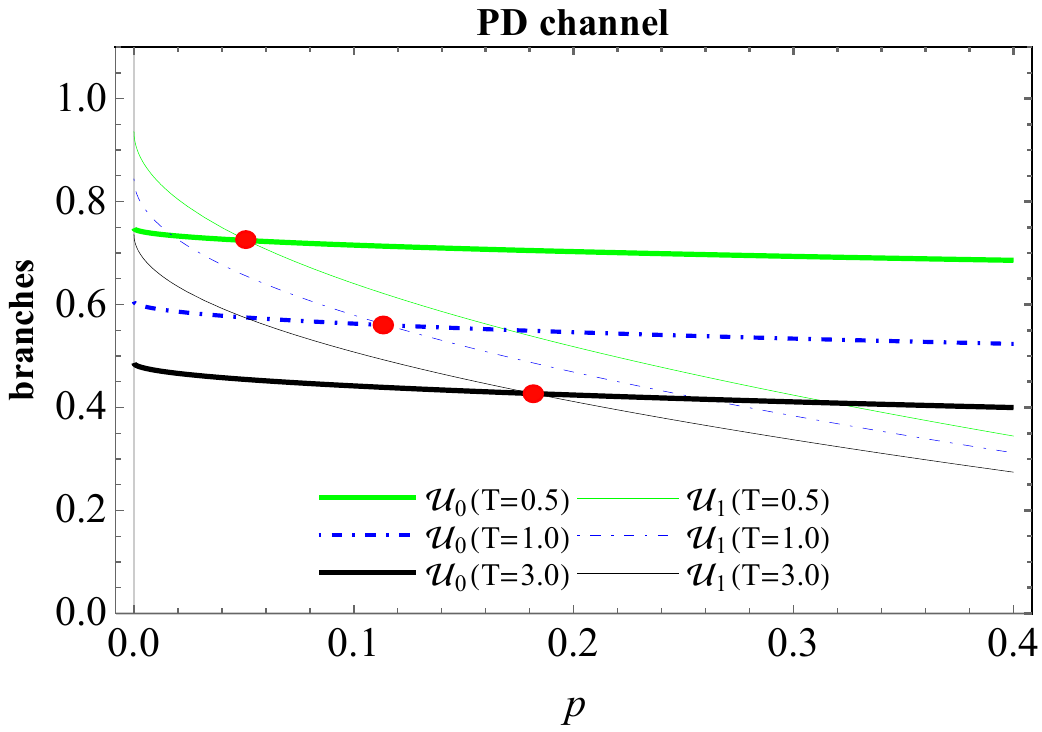}
		\put(-200,155){(b)}\qquad
  \includegraphics[width=0.45\textwidth]{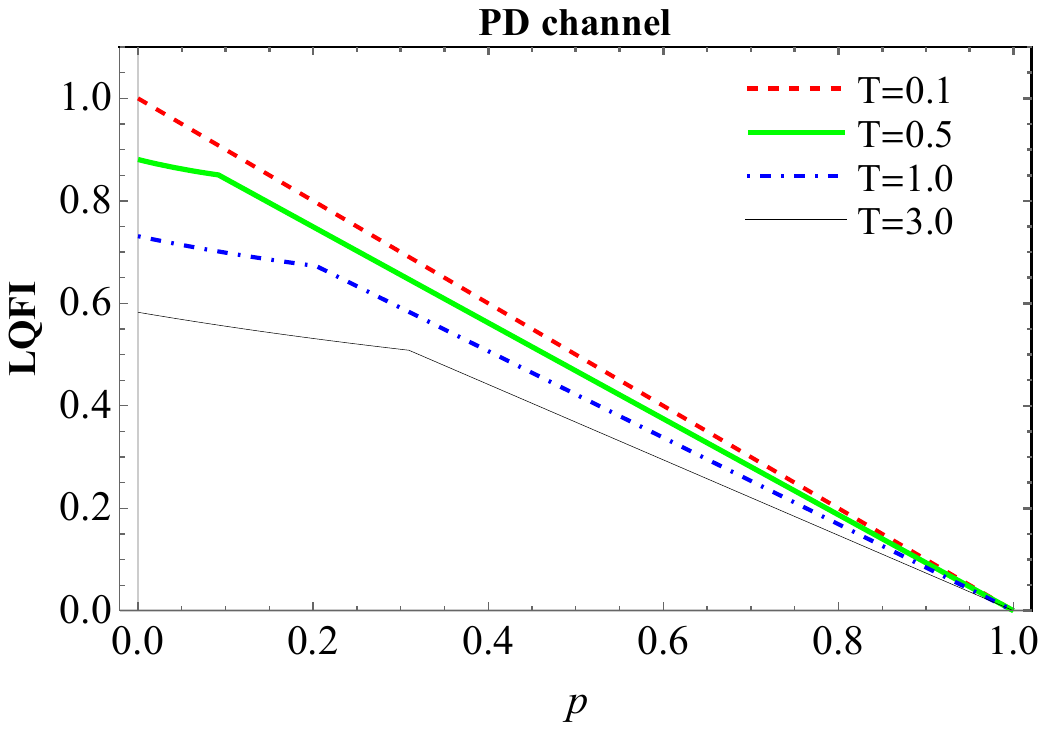}
		\put(-200,155){(c)}\quad
		\includegraphics[width=0.45\textwidth]{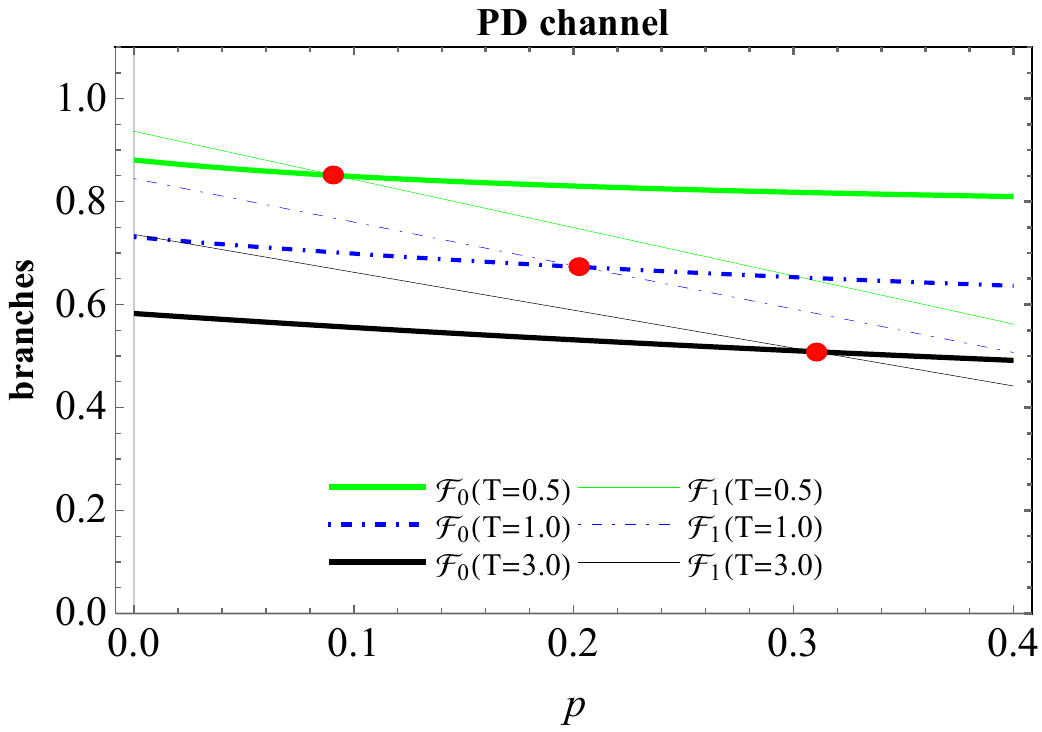}
		\put(-200,155){(d)}
	\end{center}
	\caption{LQU (a), LQFI (c), and their branches (b and d) versus $p$ when particle $A$ is exposed to PD channel and particle $B$ is positioned near the event horizon in a Schwarzschild black hole. For temperatures $T=0.5$, $T=1.0$ and $T=3.0$, the branches $\mathcal{U}_0$ and $\mathcal{U}_1$ intersect respectively at $p\approx0.05$, $p\approx0.11$ and $p\approx0.18$, and branches $\mathcal{F}_0$ and $\mathcal{F}_1$ at $p\approx0.09$, $p\approx0.20$ and $p\approx0.31$. The red bullets in plots (b) and (d) show intersection points that lead to sudden changes in the behavior of LQU and LQFI.}
	\label{figure1}
\end{figure*}

Figure \ref{figure1} delineates the quantum correlations and their branches for a non-classically correlated two-qubit state, which are quantified through the LQU and LQFI measures under the influence of Hawking radiation for particle $B$ while the particle $A$ passes through the PD channel. As evidenced in Fig. \ref{figure1}(a), the quantum correlation indicated by the function $\mathcal{U}$ \eqref{lqubranches} diminishes progressively as the severity of the decoherence parameter of the PD channel intensifies, ultimately arriving at an absence of quantum correlation. Additionally, the maximum values of quantum correlation are observed to decrease with increasing Hawking temperature. More specifically,  at $ T = 0.1 $, the LQU initiates at its maximum value ($ \mathcal{U} = 1 $) for $ p = 0 $, after which the quantum correlations decline with escalations in $ p $. Moreover, for $ T = 0.5, 1, 3 $ and at the onset of the interaction $ (p \simeq 0) $, the values of $ \mathcal{U} $  are almost $ 0.75, 0.60 $, and $ 0.49 $, respectively. There also exists a locus in these instances where an abrupt alteration in the behavior of the function transpires, wherein the decay rate intensifies.  More precisely,  the function $ \mathcal{U} $ is smooth at $T=0.1$. However, it experiences  abrupt transitions at  $p\approx0.05$, $p\approx0.11$, and $p\approx0.18$ for temperatures $T=0.5$, $T=1$, and $T=3$, respectively. As seen in Fig. \ref{figure1}(b), the branches at these points intersect and change from $\mathcal{U}_0$ to $\mathcal{U}_1$, implying sudden transitions for LQU \cite{Yurischev2023,haddadi2023aej}.

Figure  \ref{figure1}(c) elucidates that the function $ \mathcal{F} $ \eqref{lqfibranches} under analogous conditions exhibits comparable behavior to the function $ \mathcal{U} $. The quantum correlation, as quantified by $ \mathcal{F} $, decreases as $ p $ increases. However, the decay rate of $ \mathcal{F} $ is inferior to that of $ \mathcal{U} $. Moreover, at discrete $ T $ values, the function $ \mathcal{F} $ initiates at $ 1, 0.88, 0.73, $ and $ 0.58 $ for $ T = 0.1, 0.5, 1, $ and $ 3 $, respectively. Notice that the abrupt transition points for LQFI are less pronounced and shift toward escalating $ p $.   At $T=0.1$, the function  $\mathcal{F}$ shows smooth behavior similar to $\mathcal{U}$. Nonetheless, the branches of $\mathcal{F}$, namely $\mathcal{F}_0$ and $\mathcal{F}_1$, intersect at points $p\approx0.09$, $p\approx0.20$ and $p\approx0.31$ for temperatures $T=0.5$, $T=1$, and $T=3$, respectively [see Fig. \ref{figure1}(d)]. As a result, LQFI undergoes sudden transitions at these points \cite{Yurischev2023}.
Overall, at $ p = 1 $, the quantum correlations estimated by both functions are dissipated.

\begin{figure*}[t]
	\begin{center}
		\includegraphics[width=0.45\textwidth]{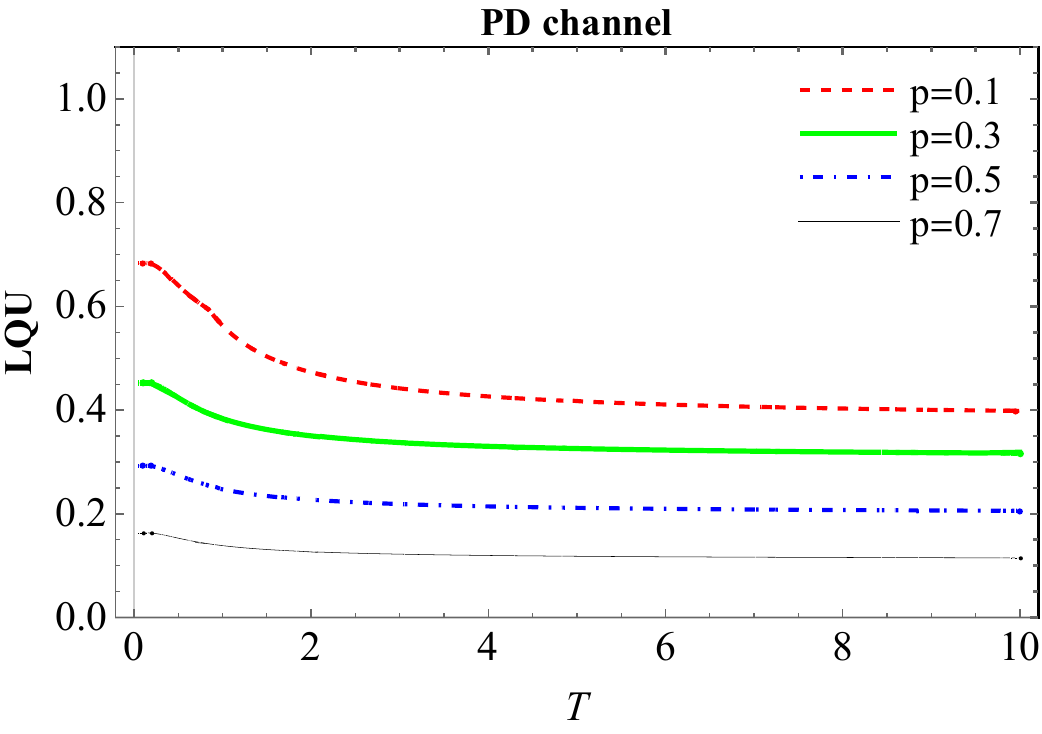}
		\put(-200,153){(a)}\qquad	
		\includegraphics[width=0.45\textwidth]{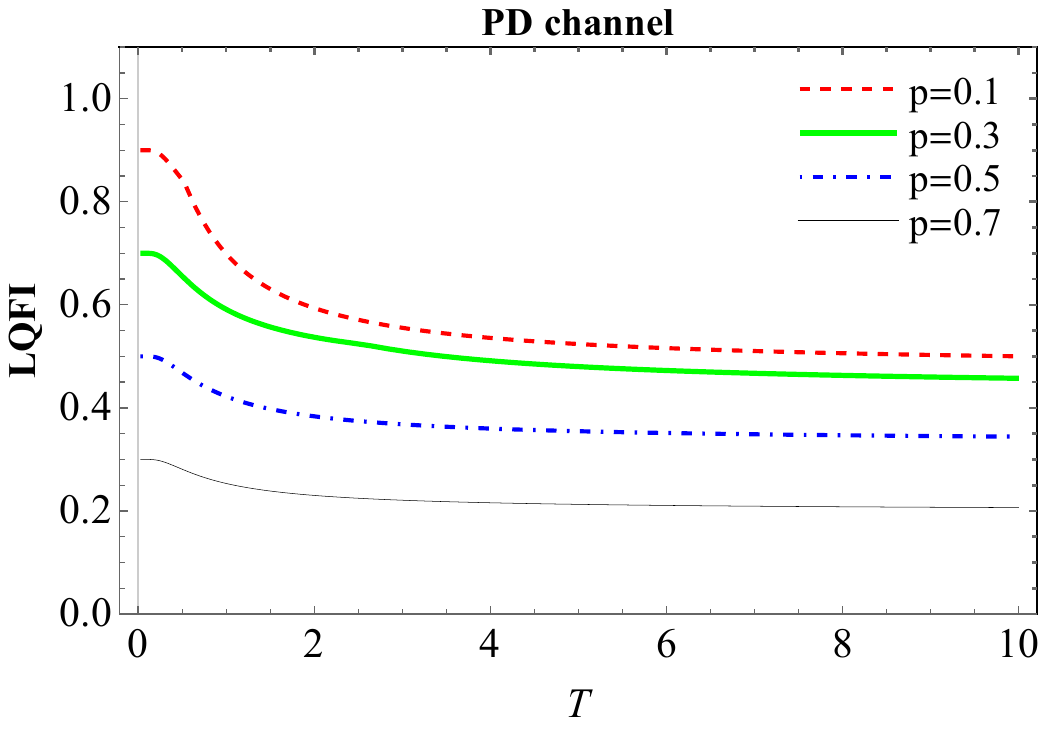}
		\put(-200,153){(b)}
	\end{center}
	\caption{LQU (a) and LQFI (b) versus $T$ when particle $A$ is exposed to PD channel and particle $B$ is positioned near the event horizon in a Schwarzschild black hole.}
	\label{figure2}
\end{figure*}

\begin{figure*}[t]
	\begin{center}
		\includegraphics[width=0.45\textwidth]{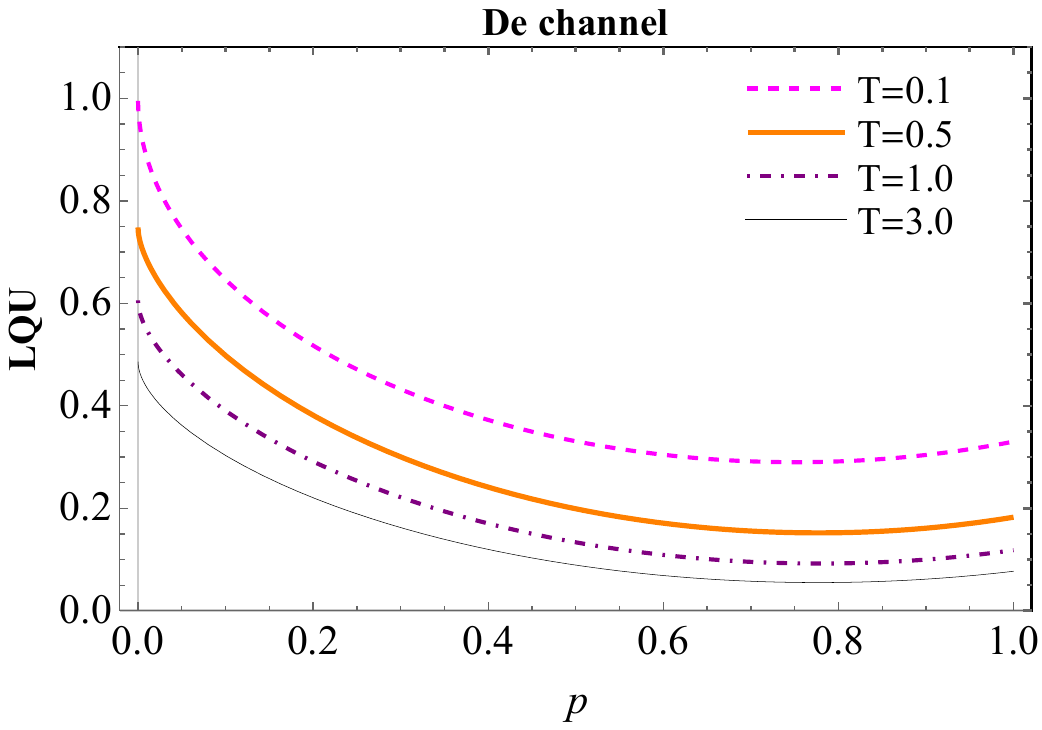}
		\put(-200,153){(a)}\qquad	
		\includegraphics[width=0.45\textwidth]{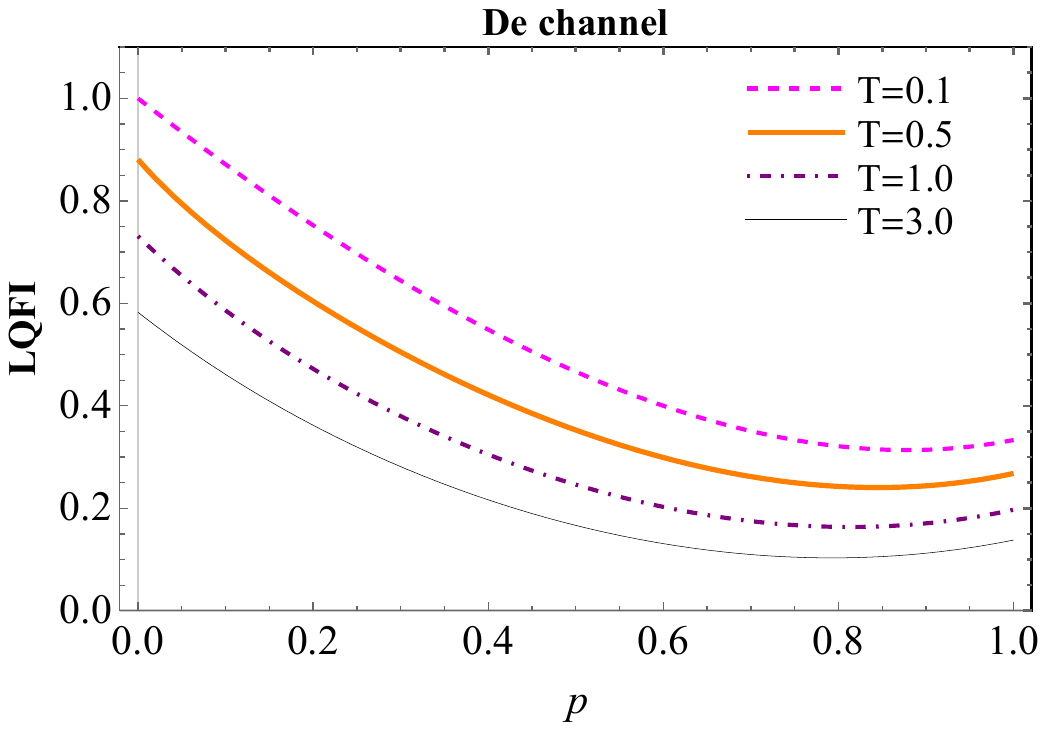}
		\put(-200,153){(b)}
	\end{center}
	\caption{LQU (a) and LQFI (b) versus $p$ when particle $A$ is undergoing De channel and particle $B$ is positioned near the event horizon in a Schwarzschild black hole.}
	\label{figure3}
\end{figure*}

In Fig \ref{figure2}, we examine the quantum correlations of the two functions $\mathcal{U}$ and $\mathcal{F}$ across continuous values of the Hawking temperature  $ T $ and discrete values of the PD channel decoherence parameter $ p $. The first qubit, $A$, passes through the PD channel, whilst the second qubit, $B$, is impacted by Hawking radiation. The outcomes depicted in Figs. \ref{figure2}(a) and  \ref{figure2}(b) demonstrate that LQU and LQFI do not attain their maximal values for any value of $ p $. Additionally, the behaviors of the functions are not considerably contingent on elevated temperatures. Intensifying $ p $ increases the decoherence of the system, as the amounts of $\mathcal{U}$ and $ \mathcal{F} $ at $ p = 0.1 $ exceed that at $ p = 0.7 $. Moreover, in agreement with Fig. \ref{figure1}, incrementing $ p $ attenuates the temperature dependence.
The extent of quantum correlations changes with $ p $ values, as the maximum bounds of quantum correlations diminish, and discrepancies between the curves become evident as decoherence strength intensifies.

\begin{figure*}[t]
	\begin{center}
		\includegraphics[width=0.45\textwidth]{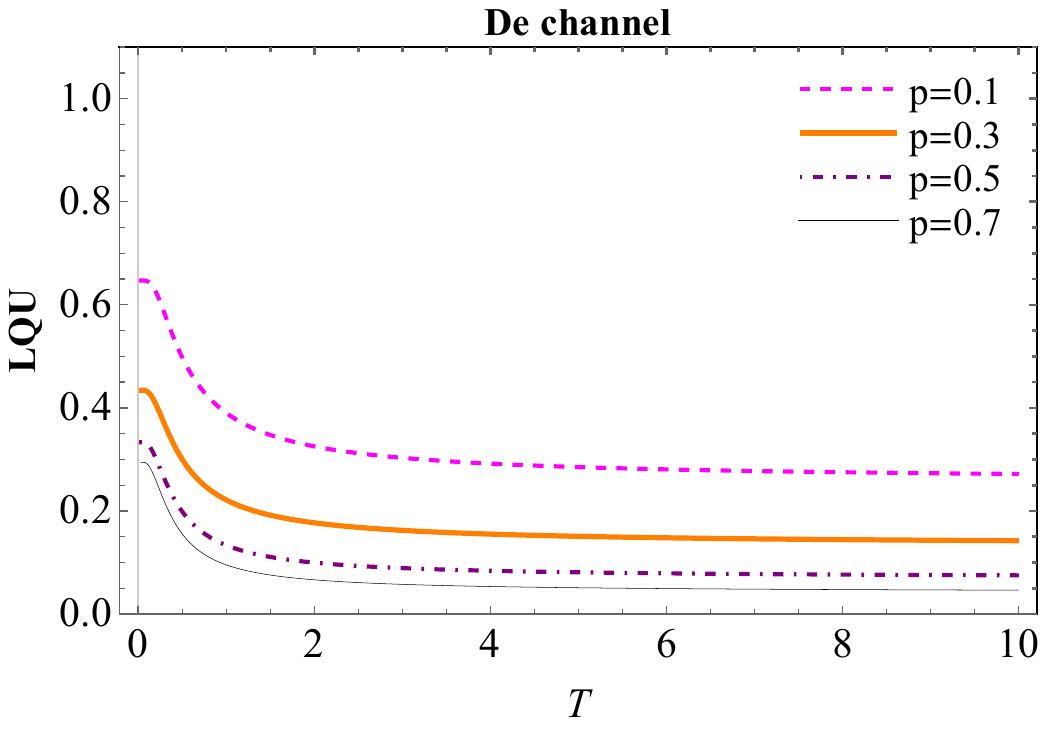}
		\put(-200,153){(a)}\qquad	
		\includegraphics[width=0.45\textwidth]{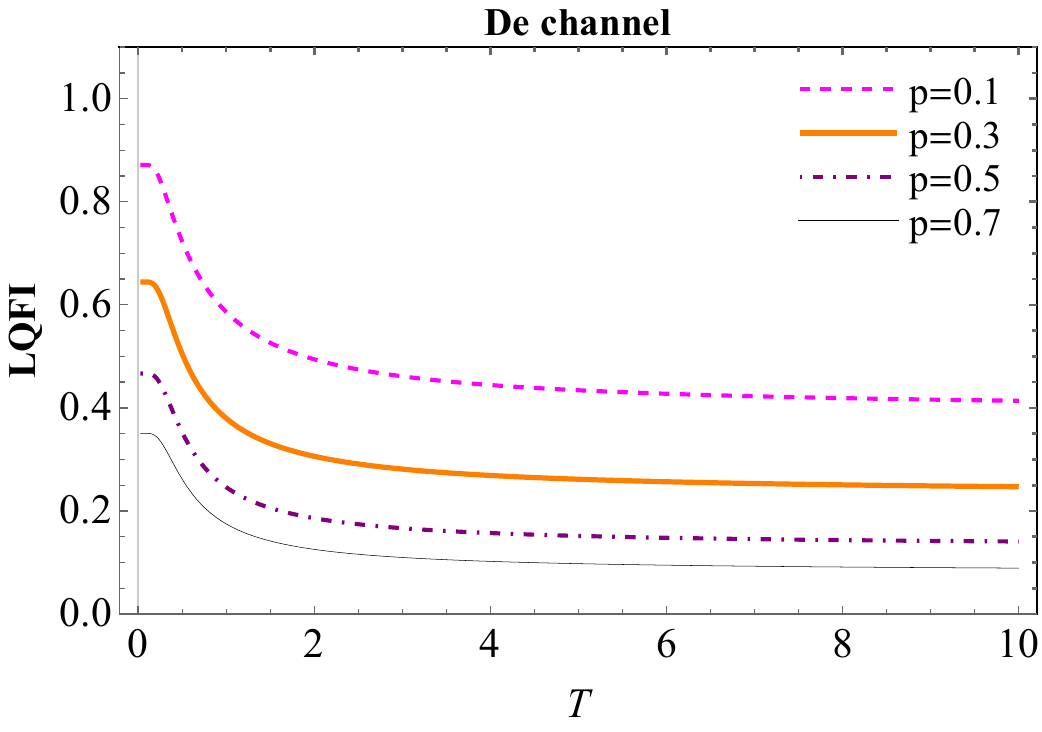}
		\put(-200,153){(b)}
	\end{center}
	\caption{LQU (a) and LQFI (b) versus $T$ when particle $A$ is undergoing De channel and particle $B$ is positioned near the event horizon in a Schwarzschild black hole.}
	\label{figure4}
\end{figure*}

The numerical outcomes of the functions LQU \eqref{lquDe} and LQFI \eqref{lqfiDe} as a function of the De decoherence parameter are delineated in Fig. \ref{figure3}, accounting for discrete Hawking temperature values. It is discernible that the two functions $ \mathcal{U} $ and $\mathcal{F}$ decay monotonically toward a non-zero steady-state value, diminishing as the decoherence parameter $ p $ escalates, before re-surging to augment toward a residual value at superior $ p $. Fig. \ref{figure3}(a) demonstrates the non-zero steady-state value is contingent on temperature; for $ T = 0.1, 0.3, 1, $ and $ 3 $ at $ p = 1 $, the residual $\mathcal{U}$ values are almost $ 0.33, 0.18, 0.12, $ and $ 0.08 $, respectively. Besides, Fig. \ref{figure3}(b) elucidates the residual value of LQFI under analogous conditions as $\mathcal{F}$ $ = 0.33, 0.27, 0.20, $ and $ 0.14 $, respectively.

In Fig. \ref{figure4}, the universal behaviors of LQU and LQFI as a function of the Hawking temperature $ T $ across discrete values of the De decoherence parameter $ p $ are displayed. Fig. \ref{figure4}(a) illustrates the behavior of LQU, which initially decreases from its initial values and then stabilizes at a specific value corresponding to the given decoherence parameter. For high temperatures, the function $\mathcal{U}$ becomes a constant with a fixed value that varies according to the value of the decoherence parameter $ p $. Similarly, Fig. \ref{figure4}(b) demonstrates that LQFI exhibits analogous behavior to LQU, but with higher upper bounds.
In general, although the quantitative behavior of LQU and LQFI is different as expected, their qualitative behavior is the same.

\begin{figure*}[t]
	\begin{center}
		\includegraphics[width=0.45\textwidth]{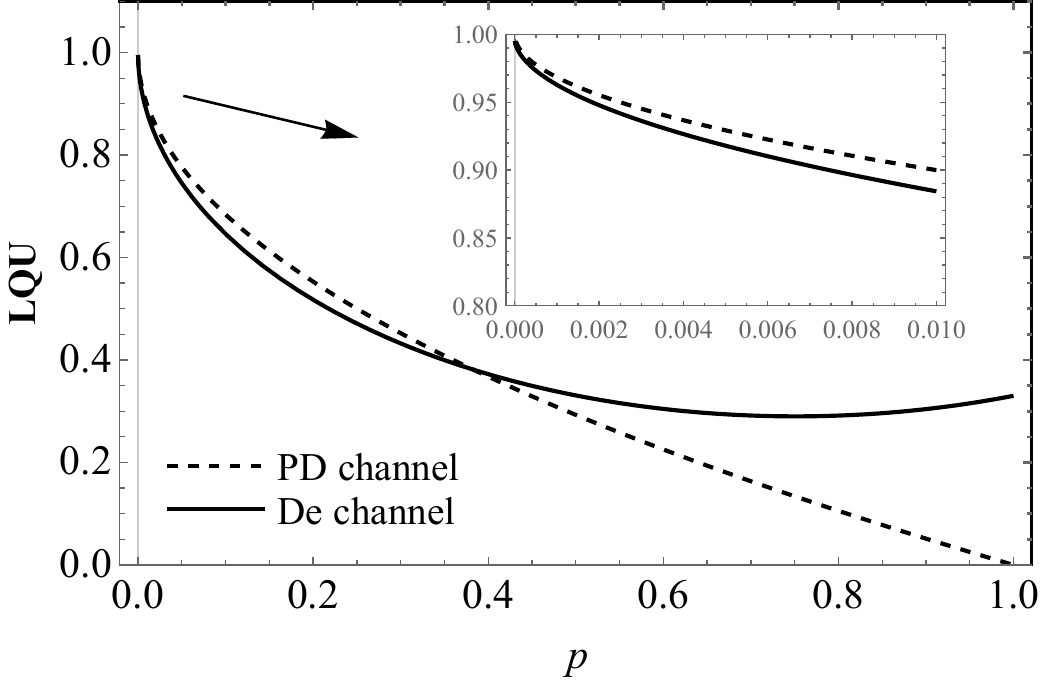}
		\put(-200,153){(a)}\qquad	
		\includegraphics[width=0.45\textwidth]{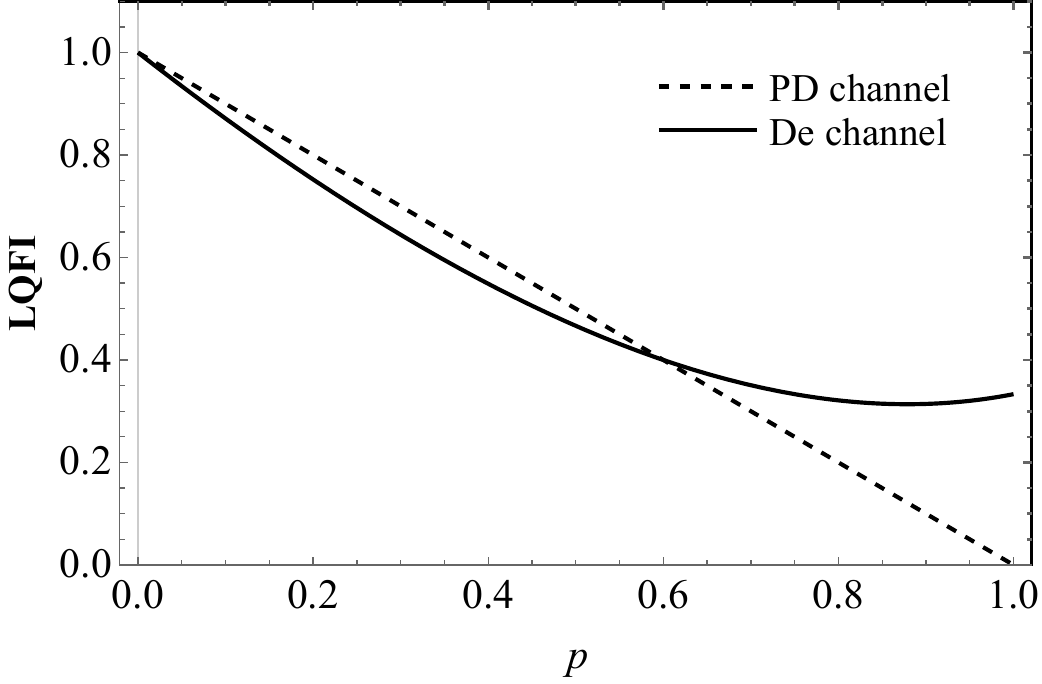}
		\put(-200,153){(b)}
	\end{center}
	\caption{LQU (a) and LQFI (b) versus $p$ when particle $A$ is exposed to PD and De channels independently and particle $B$ is positioned near the event horizon in a Schwarzschild black hole with $T=0.1$.}
	\label{figure5}
\end{figure*}

Figure \ref{figure5} presents a comparative study of LQU and LQFI functions by considering the effects of Hawking radiation and decohering channels as a function of $p$ at the fixed value of Hawking temperature $T=0.1$. Figure \ref{figure5}(a) illustrates how the variation of $p$, from $0$ to $1$, causes differences in the behaviors of LQU related to the impacts of PD and De channels. At low values of $p$, especially when $p$ tends to zero, the behavior of mentioned curves are coincident, but they are not totally similar. At low values of $p$, the LQU subjected to both PD and De channels is decreasing and the value of LQU under the PD channel is slightly more than that of under the De channel even when $p$ is close to zero. For a better illustration, a magnified frame of that region is presented in the inset plot. However, after their intersection and especially at high values of $p$, their behaviors are extremely different and the LQU values related to the PD channel are much lesser than the values of the De channel. In other words, after the intersection, the LQU curve extracted by considering the De channel keeps its value, approximately, but the LQU curve subjected to the PD channel tends to zero. The second diagram of Fig. \ref{figure5}, plot (b), provides a comparison of LQFI by considering the influences of PD and De channels. Similar to the first plot, there are resemblances in their behavior. They behave similarly to each other in the lower and middle regions of the $p$ interval. In Fig. \ref{figure5}(b) also there is an intersection. Just like previously described in Fig. \ref{figure5}(a), after the intersection the LQFI curve related to the PD channel tends to zero harshly and the curve related to the De channel maintains its value, nearly.
It is worth mentioning that the curves in Figs. \ref{figure3} and \ref{figure5} have local minima around $p=0.7$ to $p=0.9$.


\begin{figure*}[t]
	\begin{center}
		\includegraphics[width=0.45\textwidth]{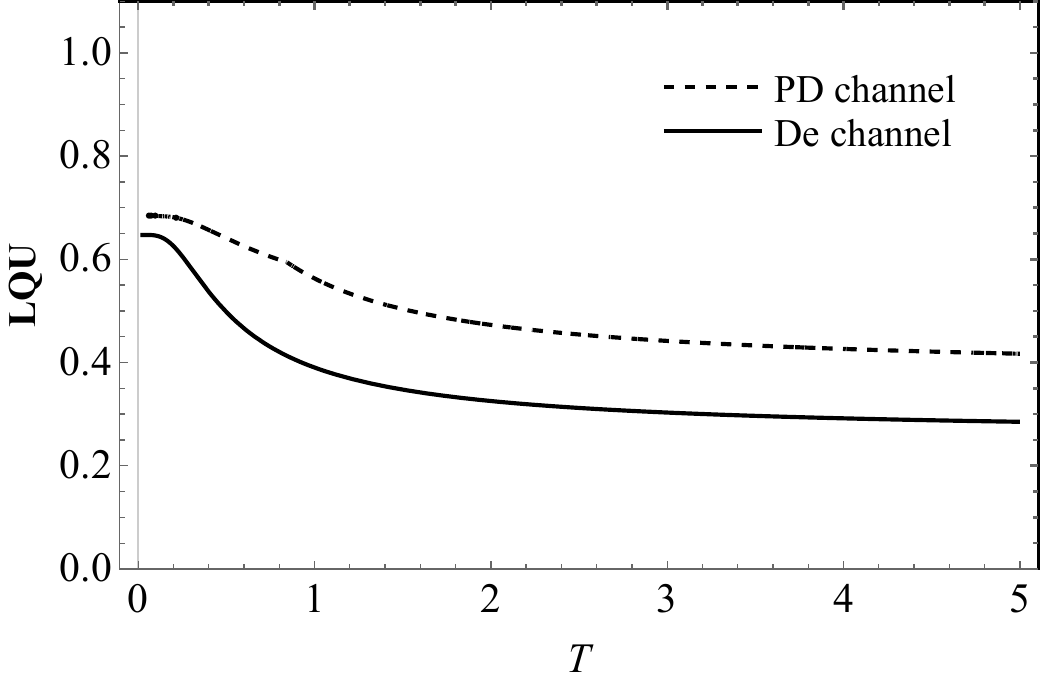}
		\put(-200,153){(a)}\qquad	
		\includegraphics[width=0.45\textwidth]{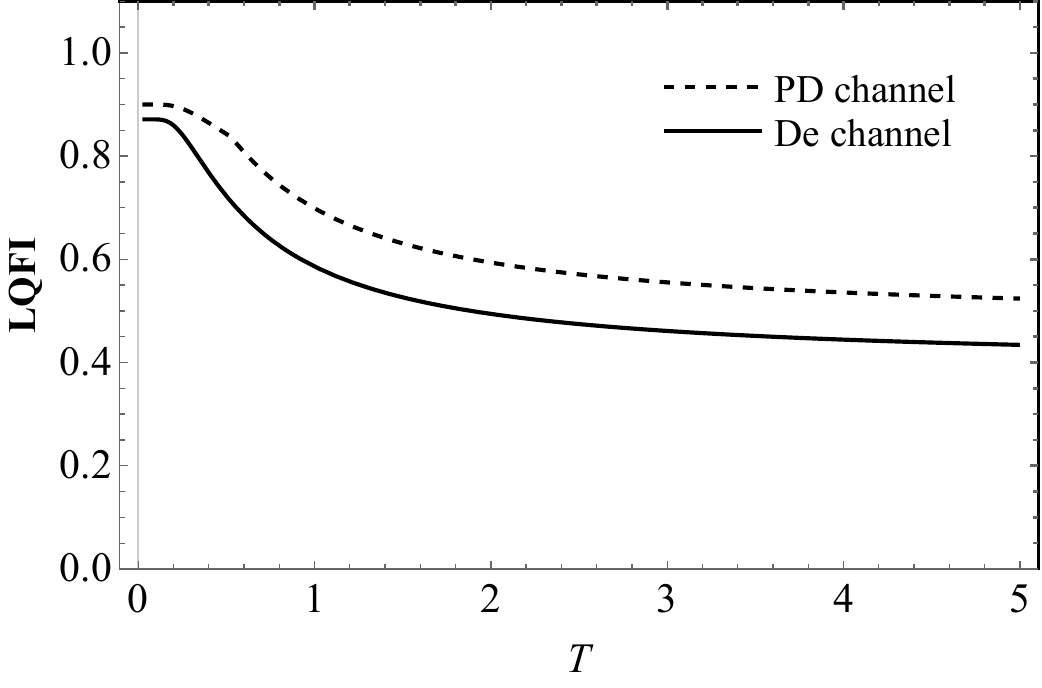}
		\put(-200,153){(b)}
	\end{center}
	\caption{LQU (a) and LQFI (b) versus $T$ when particle $A$ is exposed to PD and De channels independently and particle $B$ is positioned near the event horizon in a Schwarzschild black hole with $p=0.1$.}
	\label{figure6}
\end{figure*}

Figure \ref{figure6} just like Fig. \ref{figure5} presents a comparison of quantum correlations (LQU and LQFI) gained by considering the impacts of PD and De channels under the background of a Schwarzschild black hole, but this time they are functions of Hawking temperature $T$ that it varies from 0 to 5 and the value of $p$ is fixed to $0.1$. Both the first and second plots of Fig. \ref{figure6} during the whole interval of $T$  have equal gradients, approximately, in other words, there is no intersection between them. The values of quantum correlations subjected to the PD channel are higher than those of the De channel when $T$ varies in its interval. Note that according to Figs. \ref{figure5} and \ref{figure6}, it seems more appropriate to plot LQU and LQFI as a function of $p$ with a fixed value of $T$ to find their significant differences when exposed to PD and De channels.

\section{Quantum coherence under decohering channels}\label{sec:5}
Quantum coherence is the property of quantum systems in which the different states of the system are in a superposition, meaning they exist simultaneously \cite{Baumgratz2014}. It is essential for many quantum technologies, such as quantum computing and quantum cryptography. When a system is in a state of coherence, it can exhibit interference effects, which allow for precise control and manipulation of the system's behavior. However, coherence is fragile and can be easily disrupted by interactions with the environment, leading to a phenomenon known as decoherence. Therefore, maintaining coherence is a significant challenge in the development of practical quantum technologies.

Various techniques are at disposal for the evaluation of quantum coherence \cite{Streltsov2017}.  Among the various measures of coherence, the $l_{1}$-norm of coherence has been widely employed in the realm of quantum physics. It is expressed in the following manner
\begin{equation}\label{Quantumcoherence}
C(\varrho)=\sum_{i\neq j}\mid \varrho_{i,j}\mid.
\end{equation}

From Eq. \eqref{Quantumcoherence}, the $l_{1}$-norm of coherence value can be expressed as the summation of the absolute values of the off-diagonal elements pertaining to the selected basis \cite{mlhu2018}.

Using, as for LQU and LQFI, the system's states are given in Eqs. \eqref{statePD} and \eqref{stateDe}, we arrive at formulas for the $l_{1}$-norm of coherence under decohering channels

\begin{equation}\label{coherencePD}
C(\varrho_{A B_{\mathrm{I}}}^{\textmd{PD}})=2(|u^{+}| + |u^{-}|),
\end{equation}
and
\begin{equation}\label{coherenceDe}
C(\varrho_{A B_{\mathrm{I}}}^{\textmd{De}})=2(|\kappa^{+}| + |\kappa^{-}|).
\end{equation}

It is interesting to mention that based on Refs. \cite{Pei2012,Gebremariam2016},  the analytical expressions of quantum coherence given in Eqs. \eqref{coherencePD} and \eqref{coherenceDe} present the quantum consonance. The analytical expressions of $C(\varrho_{A B_{\mathrm{I}}}^{\textmd{PD}})$ and $C(\varrho_{A B_{\mathrm{I}}}^{\textmd{De}})$ at high and low temperatures are given in appendix \ref{Appendix B}.

Figure \ref{figure7} shows the evolution of quantum coherence between two qubits when qubit $A$ is affected by PD and De channels and qubit $B$ is in the vicinity of the event horizon in a Schwarzschild black hole. As shown in Fig. \ref{figure7}(a), quantum coherence decays with increasing parameter $p$ for both PD and De channels as expected. Because the superposition of quantum states, or interference patterns, is destroyed when those quantum systems interact with their environments. From this figure,  quantum coherence has bigger values when qubit $A$ is affected by the PD channel than the De channel for $0<p<0.74$, However, quantum coherence tends to zero more quickly when qubit $A$ is subjected to PD channel after $p=0.74$.

In Fig. \ref{figure7}(b), quantum coherence has been plotted versus Hawking temperature $T$ under the effect of PD and De channels. As depicted in this figure, quantum coherence remains unchanged for weak values of $T$ and then it decreases to get an approximately fixed value for both channels like the previous cases (see LQU and LQFI in Fig. \ref{figure6}). However, despite the previous case, quantum coherence in PD and De channels nearly has the same value of $0.67$ when $T$ tends to infinity.

\begin{figure*}[t]
	\begin{center}
		\includegraphics[width=0.45\textwidth]{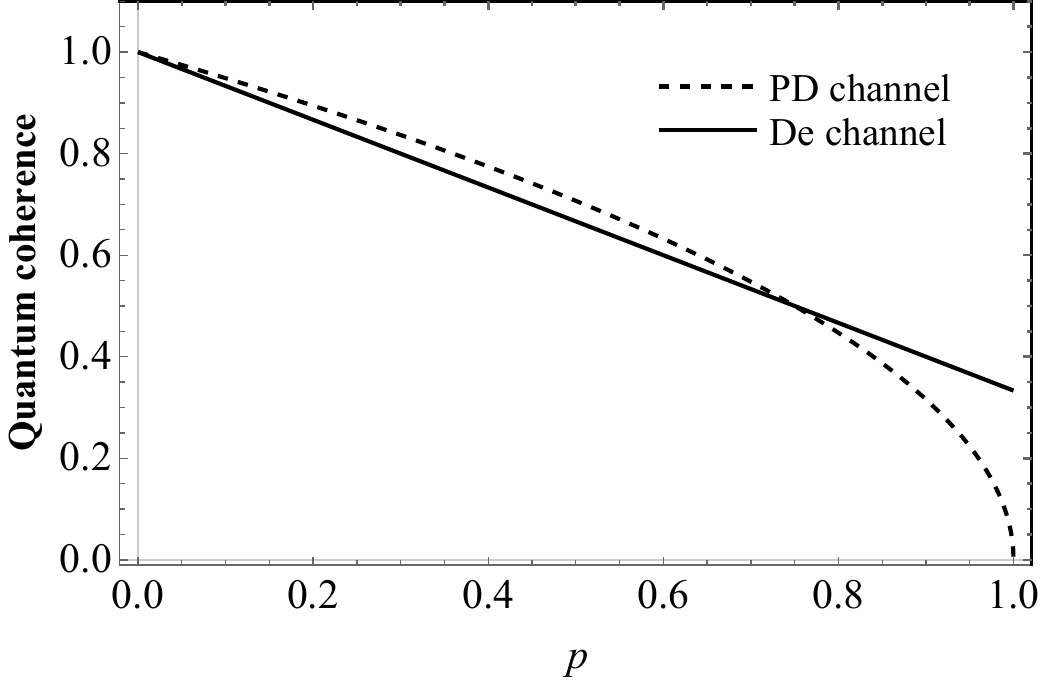}
		\put(-200,153){(a)}\quad	
		\includegraphics[width=0.45\textwidth]{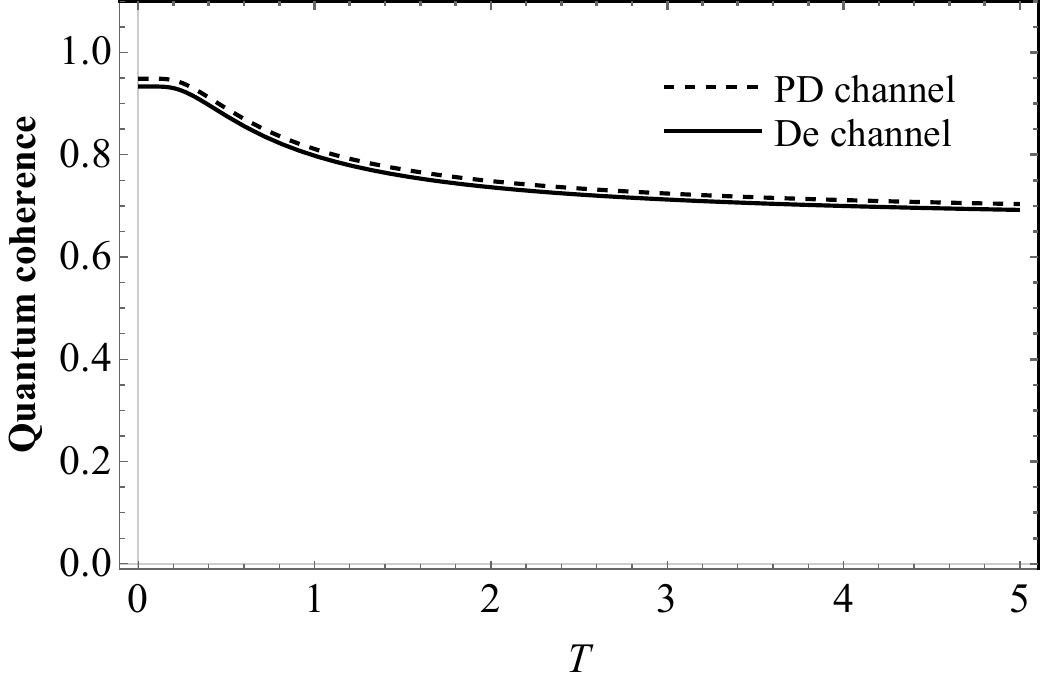}
		\put(-200,153){(b)}
	\end{center}
	\caption{Quantum coherence versus $p$ and $T$ when particle $A$ is subjected to PD and De channels independently and particle $B$ is located near the event horizon in a Schwarzschild black hole with (a) $T=0.1$ and (b) $p=0.1$.}
	\label{figure7}
\end{figure*}

\begin{figure*}[t]
	\begin{center}
		\includegraphics[width=0.45\textwidth]{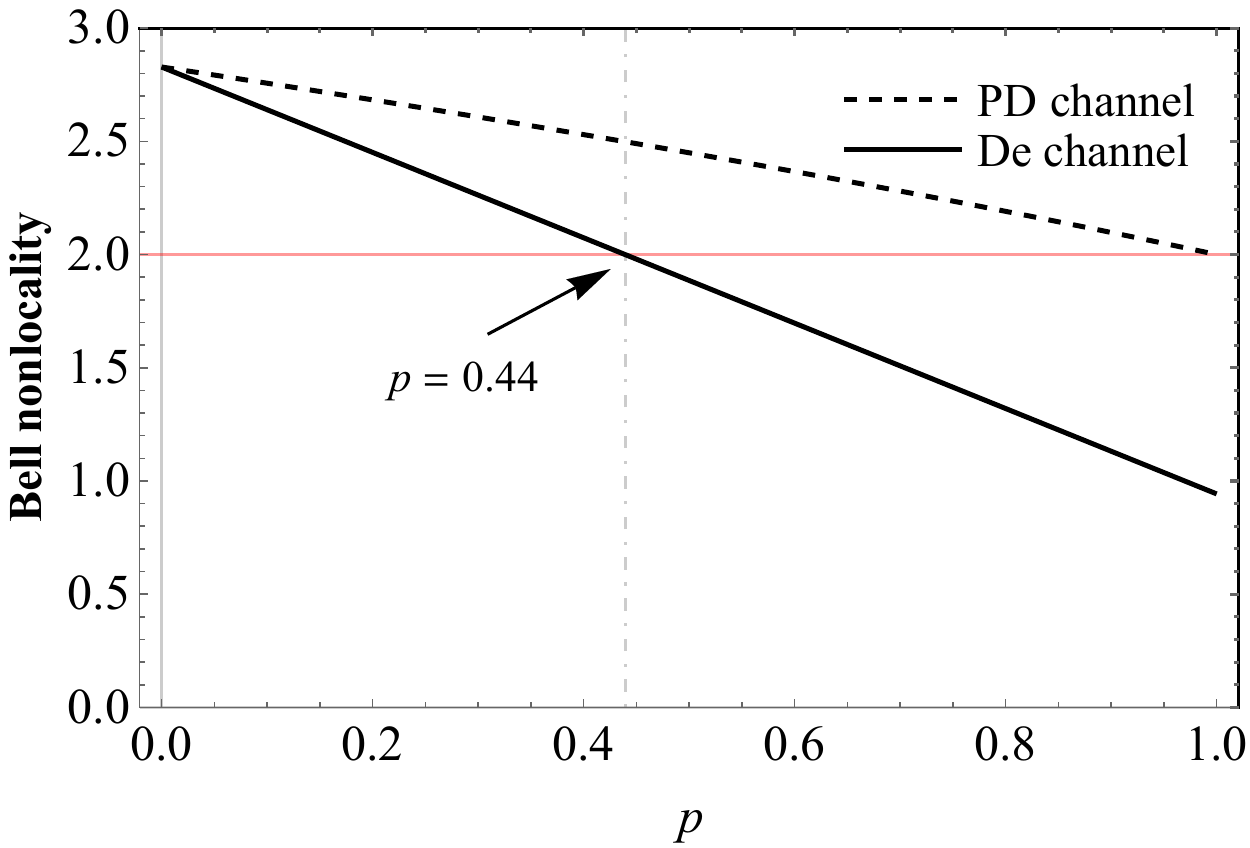}
		\put(-200,153){(a)}\qquad	
		\includegraphics[width=0.45\textwidth]{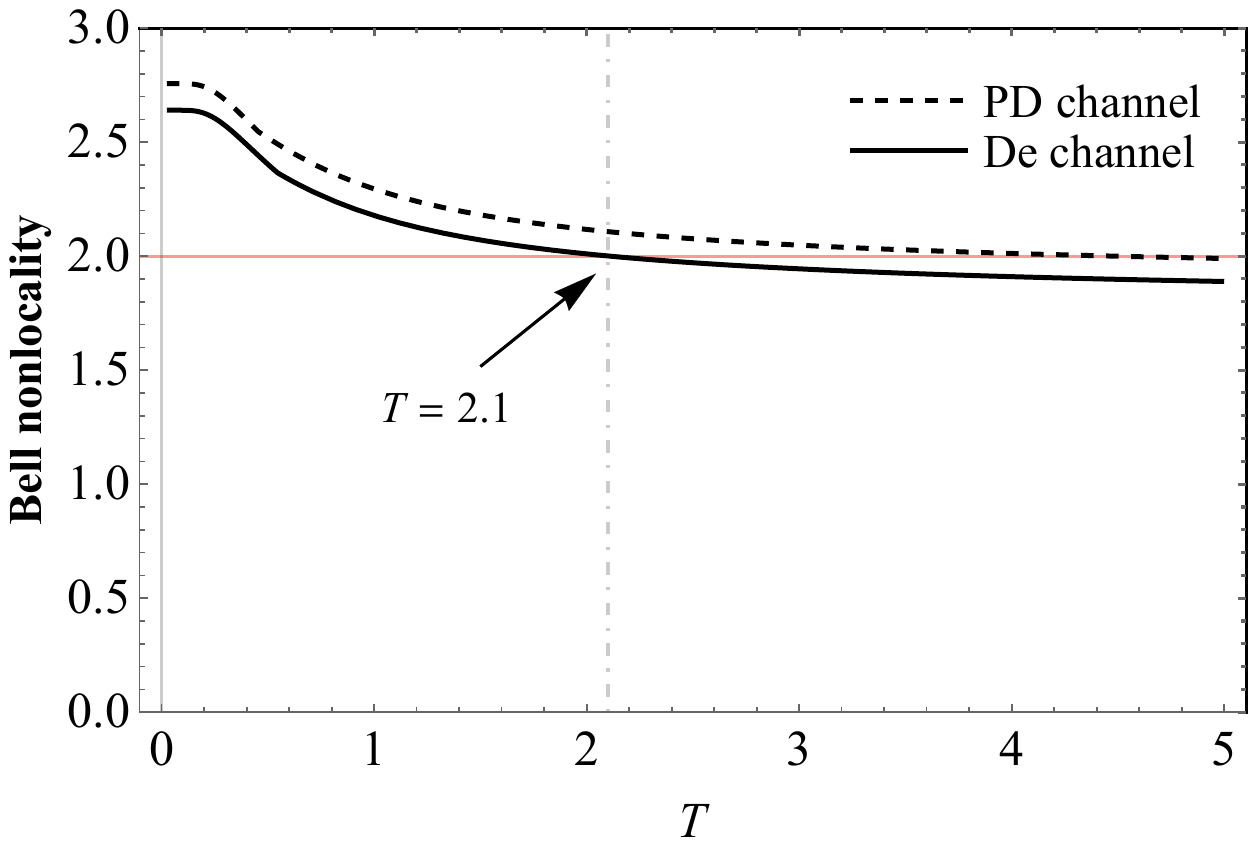}
		\put(-200,153){(b)}
	\end{center}
	\caption{Bell non-locality versus $p$ and $T$ when particle $A$ is subjected to PD and De channels independently and particle $B$ is located near the event horizon in a Schwarzschild black hole with (a) $T=0.1$ and (b) $p=0.1$.}
	\label{figure8}
\end{figure*}

\section{Bell non-locality in the Schwarzschild space-time}\label{sec:6}
Bell non-locality is a quantum phenomenon characterized by the presence of correlations between measurements performed on entangled particles that defy the predictions of classical physics \cite{Bell1964,Brunner2014}. This intriguing phenomenon strongly indicates that the behavior of the particles in question is inherently non-local and thus cannot be explained by invoking local hidden variables \cite{CHSH1969}.

For systems of dimension $2\times2$, the detection of the non-locality of a quantum state $\varrho$ can be achieved through the violation of the Bell-CHSH inequality, as proposed by Clauser, Horne, Shimony and Holt \cite{CHSH1969}. This inequality is defined as follows
\begin{equation}\label{Bell-CHSH}
|\langle B_{\textmd{CHSH}}\rangle_{\varrho}|\leq 2,
\end{equation}
with $\langle B_{\textmd{CHSH}}\rangle_{\varrho}=\textmd{tr}[\varrho B_{\textmd{CHSH}}]$ where $B_{\textmd{CHSH}}$ is the Bell operator related to the quantum CHSH inequality. The maximum value of the inequality \eqref{Bell-CHSH}, namely $B_{\max}(\varrho)=\max |\langle B_{\textmd{CHSH}}\rangle_{\varrho}|$, is related to the quantity $M(\varrho)=\max_{i<j}(\omega_i + \omega_j)$ with $B_{\max}(\varrho)=2\sqrt{M(\varrho)}$ in which $\omega_i$ ($i=1,2,3$) being the eigenvalues of a $3\times3$ matrix $X^\dagger X$, where $X$ here is a positive matrix with elements $x_{nm}=\textmd{tr}(\varrho \sigma_n \otimes \sigma_m)$ \cite{Horodecki1995}. The violation of inequality \eqref{Bell-CHSH} is manifestly evident if and only if $M(\varrho)$ exceeds the value of 1. Furthermore, $M(\varrho)$ can be utilized as an effective measure to evaluate the degree of Bell non-locality violation for a bipartite state \cite{mlhuqinp2013}.

For the quantum state expressed in Eq. \eqref{statePD}, we can obtain
\begin{equation}\label{bellpd}
B_{\max}(\varrho_{A B_{\mathrm{I}}}^{\textmd{PD}})=2\sqrt{M(\varrho_{A B_{\mathrm{I}}}^{\textmd{PD}})},
\end{equation}
where
\begin{equation}
M(\varrho_{A B_{\mathrm{I}}}^{\textmd{PD}})=\max\{M_1^{\textmd{PD}}, M_2^{\textmd{PD}}\},
\end{equation}
with
$$M_1^{\textmd{PD}}=8(|u^{+}|^2 +|u^{-}|^2),$$
and
$$M_2^{\textmd{PD}}=4(|u^{+}| +|u^{-}|)^2 + (v^{+} + \mu^{-}- v^{-} - \mu^{+})^2.$$

Also, we obtain the maximum value of the inequality \eqref{Bell-CHSH} for the state given in Eq. \eqref{stateDe} as follows
\begin{equation}\label{bellde}
B_{\max}(\varrho_{A B_{\mathrm{I}}}^{\textmd{De}})=2\sqrt{M(\varrho_{A B_{\mathrm{I}}}^{\textmd{De}})},
\end{equation}
where
\begin{equation}
M(\varrho_{A B_{\mathrm{I}}}^{\textmd{De}})=\max\{M_1^{\textmd{De}}, M_2^{\textmd{De}}\},
\end{equation}
with
$$M_1^{\textmd{De}}=8(|\kappa^{+}|^2 +|\kappa^{-}|^2),$$
and
$$M_2^{\textmd{De}}=4(|\kappa^{+}| +|\kappa^{-}|)^2 + (\vartheta^{+} + \eta^{-}- \vartheta^{-} - \eta^{+})^2.$$

Note that the analytical formulas of $B_{\max}(\varrho_{A B_{\mathrm{I}}}^{\textmd{PD}})$ and $B_{\max}(\varrho_{A B_{\mathrm{I}}}^{\textmd{De}})$ for high- and low-temperature limits are reported in appendix \ref{Appendix B}.

Figure \ref{figure8} is devoted to showing the behavior of Bell non-locality based on Eqs. \eqref{bellpd} and \eqref{bellde} when particle $A$ is exposed to PD and De channels independently and qubit $B$ is close to the event horizon in a Schwarzschild black hole. In Fig. \ref{figure8}(a), Bell non-locality has been drawn versus $p$ in which the maximum value of non-locality starts at 3 when there is no interaction between qubit $A$ and the environment, i.e., $p=0$. Bell non-locality decreases for both channels when particle $A$ begins to interact with the environment, but, it decreases for the De channel more rapidly. As shown in this figure by an arrow, Bell non-locality based on $B_{\max}(\varrho_{A B_{\mathrm{I}}}^{\textmd{De}})$ in Eq. \eqref{bellde} becomes less than 2 when $p>0.44$, however,  $B_{\max}(\varrho_{A B_{\mathrm{I}}}^{\textmd{PD}})$ in Eq. \eqref{bellpd} goes never under 2. Hence, Bell inequality is violated for all values of $p$ in the PD channel.

By looking at Figs. \ref{figure5}, \ref{figure7}(a) and \ref{figure8}(a), one can see that LQU, LQFI, and quantum coherence in our considered system when qubit $A$ is exposed to PD channel have greater values than De channel for weak values of $p$. But, Bell non-locality in the PD channel is more robust than the De channel for all values of $p$.

We have plotted Bell non-locality versus $T$ for De and PD channels in Fig. \ref{figure8}(b). It is clear that Bell non-locality has a fixed value for low Hawking temperature like previous cases in Figs. \ref{figure6} and \ref{figure7}. Then, it goes down smoothly to get values less than 2 for which non-locality effects disappear in the De channel when $T>2.1$, and we have $B_{\max}(\varrho_{A B_{\mathrm{I}}}^{\textmd{De}})\simeq1.8$ for $T\rightarrow \infty$. Nonetheless, Bell non-locality value for the PD channel becomes less than 2, i.e. $B_{\max}(\varrho_{A B_{\mathrm{I}}}^{\textmd{PD}})\simeq1.9$, when $T\rightarrow \infty$. Thus, one may infer that PD and De channels have more serious effects on quantum correlations or other characteristics of quantumness (quantum coherence and non-locality) than the effect of Hawking temperature.

\section{Conclusion and outlook}\label{sec:7}
The research conducted in the field of convergence of quantum information theory and black hole physics is promising in revealing new perspectives regarding the fundamental characteristics of gravity in the realm of quantum scales. Such research has the potential to have significant implications for our understanding of black holes as well as the complex fabric of space-time.

Motivated to investigate the relationship between these two important concepts, we considered a practical model in which a particle residing in the flat region interacts with its surroundings while the other particle experiences free fall in the vicinity of the event horizon of a Schwarzschild black hole. Based on this scenario, we studied the quantum correlations, quantum coherence, and non-locality under the collective effects of Hawking radiation and two decohering channels (PD and De).

As expected, the quantumness (quantum correlations, quantum coherence, and non-locality) of the considered system consisting of two maximally non-classically correlated qubits in the background of the Schwarzschild black hole decreases with increasing decoherence parameter $p$.  It is worthwhile to note that in the case of the PD channel,  quantum correlations and quantum coherence tend to zero when $p\to1$. However, in the case of the De channel, although they decrease, they do not go to zero. Interestingly, quantum correlations experience abrupt transitions in certain values of $p$ at fixed temperatures when qubit $A$ is exposed to the PD channel, but they decrease with increasing $p$ in a monotonic way when Alice's qubit is subjected to the De channel.

Regarding the effect of Hawking radiation, it is not an exciting result that increasing the Hawking temperature reduces quantum correlations, quantum coherence, and non-locality. But the intriguing result is that at high temperatures ($T\to \infty$), corresponding to the case of the black hole approximating to evaporate completely, we observe that the two-qubit system still preserves its quantumness, which is of great importance in quantum information theory for practical goals. Notably, when the Bell non-locality is present, its strength diminishes by enlarging the intensity of the Hawking effect.

These results not only explain how quantum correlations and quantum coherence decay but also provide better insight into quantum non-locality near black holes. Hence, our findings are helpful to guide us in choosing suitable quantum channels and quantum states to tackle relativistic quantum information processing tasks.

\vspace{20pt}

\section*{Acknowledgments}
This research is supported by the Postdoc grant of the Semnan University under Contract No. 21270. M.A.Y. was supported in part by a state task, the state registration number of the Russian Federation is FFSG-2024-0002. M.A. thanks Saeed’s Quantum Information Group (SSQIG) for providing hospitality during the course of this work.
\\
\section*{Disclosures}
The authors declare that they have no known competing financial interests.
\\
\section*{Data availability}
No datasets were generated or analyzed during the current study.
\\

\appendix
\section{LQU and LQFI at high and low temperatures}
\label{Appendix A}
In this appendix, we provide the analytical expressions of $\mathcal{W}_{11(33)}$ and $\mathcal{M}_{11(33)}$ in two cases  (high- and low-temperature limits) for PD and De channels.

From expression \eqref{eq19}, we obtain
\begin{equation}
\small
\lim_{T\to\infty}\mathcal{W}_{11}^{\textmd{PD}}(T)=\frac{2+\sqrt{2p}}{2\sqrt{3+2\sqrt{2p}}},
\end{equation}
{\small
\begin{align}
\lim_{T\to\infty}\mathcal{W}_{33}^{\textmd{PD}}(T)=\frac{(1+\sqrt{2p})^2}{3+2\sqrt{2p}},
\end{align}
and
\begin{equation}
\small
\lim_{T\to0}\mathcal{W}_{11}^{\textmd{PD}}(T)=0,
\end{equation}
{\small
\begin{align}
\lim_{T\to0}\mathcal{W}_{33}^{\textmd{PD}}(T)=\sqrt{p},
\end{align}}

Likewise, based on Eq. \eqref{eq21} we have
{\small
\begin{align}
\lim_{T\to\infty}\mathcal{M}_{11}^{\textmd{PD}}(T)= \frac{1+p}{2+p},
\end{align}}
\begin{equation}
\small
\lim_{T\to\infty}\mathcal{M}_{33}^{\textmd{PD}}(T)= \frac{1+2p}{3},
\end{equation}
and
\begin{equation}
\small
\lim_{T\to0}\mathcal{M}_{11}^{\textmd{PD}}(T)= 0,
\end{equation}

\begin{equation}
\small
\lim_{T\to0}\mathcal{M}_{33}^{\textmd{PD}}(T)= p.
\end{equation}

Further, using Eqs. \eqref{eq25} and \eqref{eq27}, we get
{\small
\begin{align}
\lim_{T\to\infty}\mathcal{W}_{11}^{\textmd{De}}(T)= &\frac{1}{144}\bigg[\frac{24 (4 p (3-5 p)+9)}{\left(w_{1-}+w_{1+}\right)\left(w_{2-}+w_{2+}\right)}\nonumber\\
&+6\left(w_{1-}+w_{1+}\right)\left(w_{2-}+w_{2+}\right)\bigg],
\end{align}}
{\small
\begin{align}
\lim_{T\to\infty}\mathcal{W}_{33}^{\textmd{De}}(T)= &\frac{1}{48}\bigg[-\frac{4 (4 p (p+3)-9)}{\left(w_{1-}+w_{1+}\right){}^2}+\left(w_{1-}+w_{1+}\right){}^2\nonumber\\
&-\frac{4 (4 p (17 p-33)+63)}{\left(w_{2-}+w_{2+}\right){}^2}+\left(w_{2-}+w_{2+}\right){}^2\bigg],
\end{align}}
where
{\small $$w_{1\pm }=\sqrt{3+2 p\pm \sqrt{3} \sqrt{4p (p-1) +3}},$$
$$w_{2\pm }=\sqrt{9-2 p\pm \sqrt{4 p (19 p-39)+81}},$$}
and
{\small
\begin{align}
\lim_{T\to0}\mathcal{W}_{11}^{\textmd{De}}(T)= \lim_{T\to0}\mathcal{W}_{33}^{\textmd{De}}(T)= \frac{2}{3} \sqrt{p(3-2 p)}.
\end{align}}

Besides
{\small
\begin{align}
\lim_{T\to\infty}\mathcal{M}_{11}^{\textmd{De}}(T)= \frac{(8 p (2 p-3)-9) (p (4 p+3)-18)}{6 [p (4 p (2 p-9)+9)+54]},
\end{align}}
{\small
\begin{align}
\lim_{T\to\infty}\mathcal{M}_{33}^{\textmd{De}}(T)=1-\frac{2}{3}\frac{[16p^3+9(3-4p)]}{(3+2p)(9-2p)},
\end{align}}
and
{\small
\begin{align}
\lim_{T\to0}\mathcal{M}_{11}^{\textmd{De}}(T)= \lim_{T\to0}\mathcal{M}_{33}^{\textmd{De}}(T)= \frac{4p(3-2p)}{3(3-p)}.
\end{align}}

\section{Quantum coherence and Bell non-locality at high and low temperatures}
\label{Appendix B}
Here, we give the analytical expressions of $C(\varrho)$ and $B_{\max}(\varrho)$ in two cases  (high- and low-temperature limits) for PD and De channels.

From expressions \eqref{coherencePD} and \eqref{coherenceDe}, we get
\begin{equation}
\small
\lim_{T\to\infty}C[\varrho_{A B_{\mathrm{I}}}^{\textmd{PD}}](T)=\sqrt{\frac{1- p}{2}},
\end{equation}
\begin{equation}
\small
\lim_{T\to0}C[\varrho_{A B_{\mathrm{I}}}^{\textmd{PD}}](T)=\sqrt{1- p},
\end{equation}
and
\begin{equation}
\small
\lim_{T\to\infty}C[\varrho_{A B_{\mathrm{I}}}^{\textmd{De}}](T)=\frac{3-2p}{3 \sqrt{2}},
\end{equation}

\begin{equation}
\small
\lim_{T\to0}C[\varrho_{A B_{\mathrm{I}}}^{\textmd{De}}](T)=\frac{3-2p}{3}.
\end{equation}

Moreover, according to Eqs. \eqref{bellpd} and \eqref{bellde} we obtain
\begin{equation}
\small
\lim_{T\to\infty}B_{\max}[\varrho_{A B_{\mathrm{I}}}^{\textmd{PD}}](T)=2 \sqrt{1-p},
\end{equation}
\begin{equation}
\small
\lim_{T\to0}B_{\max}[\varrho_{A B_{\mathrm{I}}}^{\textmd{PD}}](T)=2 \sqrt{2-p},
\end{equation}
and
\begin{equation}
\small
\lim_{T\to\infty}B_{\max}[\varrho_{A B_{\mathrm{I}}}^{\textmd{De}}](T)=\frac{2}{3} \sqrt{2 p (5 p-9)+9},
\end{equation}
\begin{equation}
\small
\lim_{T\to0}B_{\max}[\varrho_{A B_{\mathrm{I}}}^{\textmd{De}}](T)=\frac{2}{3} \sqrt{2} (3-2 p).
\end{equation}



\end{document}